\documentclass[showpacs,twocolumn,aps,prx,floatfix,superscriptaddress,noshowpacs,longbibliography]{revtex4-1}

\usepackage{dsfont}
\usepackage{amsmath,amssymb,graphicx,bm,color,mathrsfs,verbatim,epstopdf,dcolumn}

\usepackage{caption}
\usepackage{subcaption}
\graphicspath{{figs/}}

\usepackage{hyperref}
\hypersetup{ 
	colorlinks   = true
}

\newcommand*{\red}{\textcolor{red}}

\DeclareMathOperator*{\argmin}{arg\,min}
\DeclareMathOperator*{\argmax}{arg\,max}

\begin{document}

\title{Reinforcement Learning in Different Phases of Quantum Control}

\author{Marin Bukov}
\email{mbukov@bu.edu}
\affiliation{Department of Physics, Boston University, 590 Commonwealth Ave., Boston, MA 02215, USA}

\author{Alexandre G.R. Day}
\email{agrday@bu.edu}
\affiliation{Department of Physics, Boston University, 590 Commonwealth Ave., Boston, MA 02215, USA}

\author{Dries Sels}
\affiliation{Department of Physics, Boston University, 590 Commonwealth Ave., Boston, MA 02215, USA}
\affiliation{Theory of quantum and complex systems, Universiteit Antwerpen, B-2610 Antwerpen, Belgium}

\author{Phillip Weinberg}
\affiliation{Department of Physics, Boston University, 590 Commonwealth Ave., Boston, MA 02215, USA}

\author{Anatoli Polkovnikov}
\affiliation{Department of Physics, Boston University, 590 Commonwealth Ave., Boston, MA 02215, USA}

\author{Pankaj Mehta}
\affiliation{Department of Physics, Boston University, 590 Commonwealth Ave., Boston, MA 02215, USA}

\begin{abstract}
The ability to prepare a physical system in a desired quantum state is central to many areas of physics such as nuclear magnetic resonance, cold atoms, and quantum computing. Yet, preparing states quickly and with high fidelity remains a formidable challenge. In this work we implement cutting-edge Reinforcement Learning (RL) techniques and show that their performance is comparable to optimal control methods in the task of finding short, high-fidelity driving protocol from an initial to a target state in non-integrable many-body quantum systems of interacting qubits. RL methods learn about the underlying physical system solely through a single scalar reward (the fidelity of the resulting state) calculated from numerical simulations of the physical system. We further show that quantum state manipulation, viewed as an optimization problem,  exhibits a spin-glass-like phase transition in the space of protocols as a function of the protocol duration. Our RL-aided approach helps identify variational protocols with nearly optimal fidelity, even in the glassy phase, where optimal state manipulation is exponentially hard. This study highlights the potential usefulness of RL for applications in out-of-equilibrium quantum physics.
\end{abstract} 
\date{\today}



\maketitle

\section{\label{sec:intro}Introduction}

Reliable quantum state manipulation is essential for many areas of physics ranging from nuclear magnetic resonance experiments~\cite{vandersypen_05} and cold atomic systems~\cite{vanfrank_16,wigley_16} to trapped ions~\cite{islam_11,senko_15,jurcevic_14}, quantum optics~\cite{sayrin_11}, superconducting qubits~\cite{barends_16}, nitrogen vacancy centers~\cite{zhou_17}, and quantum computing~\cite{nielsen}. However,  finding optimal control sequences in such experimental platforms presents a formidable challenge due to our limited theoretical understanding of nonequilibrium quantum systems, and the intrinsic complexity of simulating large quantum many-body systems.

For long protocol durations, adiabatic evolution can be used to robustly reach target quantum states, provided the change in the Hamiltonian is slow compared to the minimum energy gap. 
Unfortunately, this assumption is often violated in real-life applications. Typical experiments often have stringent constraints on control parameters, such as a maximum magnetic-field strength or a maximal switching frequency. Moreover, decoherence phenomena impose insurmountable time constraints beyond which quantum information is lost irreversibly. For this reason, many experimentally-relevant systems are in practice uncontrollable, i.e.~there are no finite-duration protocols, which prepare the desired state with unit fidelity. In fact, in Anderson and many-body localized, or periodically-driven systems, which are naturally away from equilibrium, the adiabatic limit does not even exist~\cite{khemani_15,pweinberg_15}. This has motivated numerous approaches to quantum state control~\cite{baksic_16,wang_17,agundez_17,bao_17,rotskoff_17,leung_16,yang_16,jarzynski_13,kolodrubetz_16,sels_16,glaser_98,rabitz_00,khaneja_01,sklarz_02,demirplak_03,khaneja_05,berry_09,caneva_11,zahedinejad_14,zahedinejad_15,theisen_17,bukov_GSL,wu_2018}. 
Despite all advances, at present date surprisingly little is known about how to successfully load a non-integrable interacting quantum system into a desired target state, especially in short times, or even when this is feasible in the first place~\cite{caneva_11,doria_11,caneva_14,lloyd_14}. 

\begin{figure}[t!]
	\includegraphics[width=1.0\columnwidth]{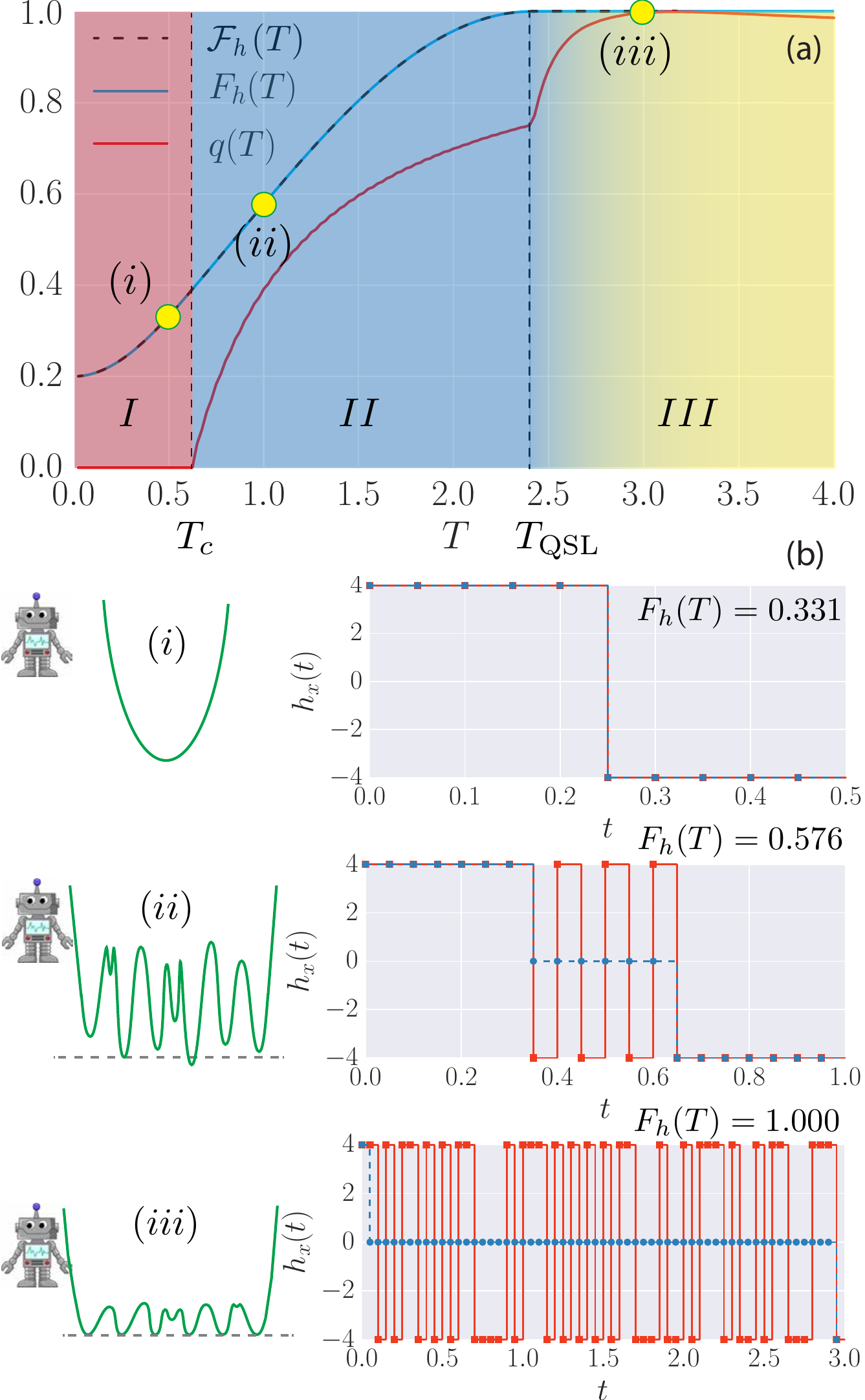}
	\caption{\label{fig:phase_diag} (a) Phase diagram of the quantum state manipulation problem for the qubit in Eq.~\eqref{eq:H_2LS} vs.~protocol duration $T$, as determined by the order parameter $q(T)$ (red) and the maximum possible achievable fidelity $F_h(T)$ (blue), compared to the variational fidelity $\mathcal{F}_h(T)$ (black, dashed). Increasing the total protocol time $T$, we go from an overconstrained phase $I$, through a glassy phase $II$, to a controllable phase $III$. (b) Left: the infidelity landscape is shown schematically (green). Right: the optimal bang-bang protocol found by the RL agent at the points (i)--(iii) (red) and the variational protocol~\cite{supplementary} (blue, dashed).   }
\end{figure}

In this paper, we adopt a radically different approach to this problem based on machine learning (ML)~\cite{sutton_barto,bishop,mnih_15_ATARI,silver_16_ALPHAGO,dunjko_16,ML_review,judson_92}. ML has recently been applied successfully to several problems in equilibrium condensed matter physics~\cite{carleo_17,carrasquilla_17}, turbulent dynamics~\cite{reddy_16,colabrese_17} and experimental design~\cite{krenn_16,melnikov_17}, and here we demonstrate that Reinforcement Learning (RL) provides deep insights into nonequilibrium quantum dynamics~\cite{chen_14_ML,august_18,foesel_18,zhang2018automatic,niu2018universal,albarran2018measurement}. Specifically, we use a modified version of the Watkins Q-Learning algorithm~\cite{sutton_barto} to teach a computer agent to find driving protocols which prepare a quantum system in a target state $|\psi_*\rangle$ starting from an initial state $|\psi_i\rangle$ by controlling a time-dependent field. A far-reaching consequence of our study is the existence of phase transitions in the quantum control landscape of the generic many-body quantum control problem. The glassy nature of the prevalent phase implies that the optimal protocol is exponentially difficult to find. However, as we demonstrate, the optimal solution is unstable to local perturbations. Instead, we discover classes of RL-motivated stable suboptimal protocols\cite{gingrich_16}, the performance of which rival that of the optimal solution. Analyzing these suboptimal protocols, we construct a variational theory, which demonstrates that the behaviour of physical degrees of freedom (d.o.f.) [which scale exponentially with the system size $L$ for ergodic models] in a non-integrable many-body quantum spin chain can be effectively described by only a few variables within the variational theory. We benchmark the RL results using Stochastic Descent (SD), and compare them to optimal control methods such as CRAB~\cite{caneva_11} and (for simplicity) first-order GRAPE~\cite{grape_05} (without its quasi-Newton extensions~\cite{machnes2011comparing,agundez_17,de2011second}), see discussion in the Supplemental Material~\cite{supplementary}.

In stark contrast to most approaches to quantum optimal control, RL is a \emph{model-free} feedback-based method which could allow for the discovery of controls even when accurate models of the system are unknown, or parameters in the model are uncertain. A potential advantage of RL over traditional derivative-based optimal control approaches is the fine balance between exploitation of already obtained knowledge and exploration in uncharted parts of the control landscape. Below the quantum speed limit~\cite{hegerfeldt_13}, exploration becomes vital and offers an alternative to the prevalent paradigm of multi-starting local gradient optimizers~\cite{sorensen_16}. Unlike these methods, the RL agent progressively learns to build a model of the optimization landscape in such a way that the protocols it finds are stable to sampling noise. In this regard, RL-based approaches may be particularly well-suited to work with experimental data and do not require explicit knowledge of local gradients of the control landscape~\cite{grape_05,supplementary}. This may offer a considerable advantage in controlling realistic systems where  constructing a reliable effective model is infeasible, for example due to disorder or dislocations.

To manipulate the quantum system, our computer agent constructs piecewise-constant protocols of duration $T$ by choosing a drive protocol strength $h_x(t)$ at each time $t=j\delta t$, $j=\left\{0, 1,\cdots,T/\delta t\right\}$, with $\delta t$ the time-step size. In order to make the agent learn, it is given a reward for every protocol it constructs -- the fidelity $F_h(T)=|\langle\psi_\ast|\psi(T)\rangle|^2$ for being in the target state after time $T$ following the protocol $h_x(t)$ \emph{under unitary Schr\"odinger evolution}. The goal of the agent is to maximize the reward in a series of attempts. Deprived of any knowledge about the underlying physical model, the agent collects information about already tried protocols, based on which it constructs new, improved protocols through a sophisticated biased sampling algorithm. In realistic applications, one does not have access to infinite control fields; for this reason, we restrict to fields $h_x(t)\in[-4,4]$, see Fig.~\ref{fig:phase_diag}b. 
For reasons relating to the simplicity and efficiency of the numerical simulations, throughout this work we further restrict the RL algorithm to the family of bang-bang protocols \footnote{We note that there is no proof that the family of bang-bang protocols contains the optimal protocol for the single-qubit control problem in Sec.~\ref{sec:phases}, since the present control problem is of bilinear type~\cite{OC_book} }. An additional advantage of focusing on  bang-bang protocols is that this allows us to interpret the control phase transitions we find using the language of Statistical Mechanics~\cite{day2018glassy}.

\section{\label{sec:RL}Reinforcement Learning}

Reinforcement Learning (RL) is a subfield of Machine Learning (ML) in which a computer agent learns to perform and master a specific task by exerting a series of actions in order to maximize a reward function, as a result of interaction with its environment. Here, we use a modified version of Watkins online, off-policy Q-Learning algorithm with linear function approximation and eligibility traces~\cite{sutton_barto} to teach our RL agent to find protocols of optimal fidelity. Let us we briefly summarize the details of the procedure. For a detailed description of the standard Q-learning algorithm, we refer the reader to Ref.~\cite{sutton_barto}.

The fidelity optimization problem is defined as an episodic, undiscounted Reinforcement Learning task. Each episode takes a fixed number of steps $N_T = T/\delta t$, where $T$ is the total protocol duration, and $\delta t$ -- the physical (protocol) time step. We define the state $\mathcal{S}$, action $\mathcal{A}$ and reward $\mathcal{R}$ spaces, respectively, as
\begin{equation*}
\mathcal{S} \!=\! \{s\!=\!(t, h_x(t))\},\quad \mathcal{A}\!=\!\{ a\!=\!\delta h_x \},\quad \mathcal{R} \!=\! \{r\in [0,1]\}.
\end{equation*}
The state space $\mathcal{S}$ consists of all tuples $(t, h_x(t))$ of time $t$ and the corresponding magnetic field $h_x(t)$. Notice that with this choice no information about the physical quantum state whatsoever is encoded in the RL state, and hence the RL algorithm is model-free. Thus, the RL agent will be able to learn circumventing the difficulties associated with the theoretical notions in quantum physics. Including time $t$ to the state is not common in Q-Learning, but is required here in order for the agent to be able to estimate how far away it is from the episode's end, and adjust its actions accordingly. Even though there is only one control field, the space of available protocols grows exponentially with the inverse step size $\delta t^{-1}$.

The action space $\mathcal{A}$ consists of all jumps $\delta h_x$ in the protocol $h_x(t)$. Thus, protocols are constructed as piecewise-constant functions. We restrict the available actions of the RL agent in every state $s$ such that at all times the field $h_x(t)$ is in the interval $[-4,4]$. We verify that RL also works for quasi-continuous protocols with many different steps $\delta h_x$~\cite{supplementary}. The bang-bang protocols discussed in the next section and the quasi-continuous protocols, used in the Supplemental Material~\cite{supplementary}, are examples of the family of protocol functions we allow in the simulation. 

Last but not least, the reward space $\mathcal{R}$ is the space of all real numbers in the interval $[0,1]$. The rewards for the agent are given only at the end of each episode, according to:
\begin{equation}
r(t) = 
\begin{cases} 
\hfill  0,    \hfill & \text{ if $t<T$} \\
\hfill  F_h(T)=|\langle\psi_\ast|\psi(T)\rangle|^2,  \hfill & \text{ if $t=T$} \\
\end{cases}
\label{eq:rewards}
\end{equation}
This reflects the fact that we are not interested in which quantum state the physical system is in during the evolution; all that matters for our purpose is to maximize the final fidelity. 

An essential part of setting up the RL problem is to define the environment, with which the agent interacts in order to learn. We choose this to consist of the Schr\"odinger initial value problem, together with the target state:
\begin{eqnarray*}
\mathrm{Environment} &=& \{ i\partial_t|\psi(t)\rangle = H(t)|\psi(t)\rangle, \nonumber\\
&& \qquad |\psi(0)\rangle=|\psi_i\rangle, \quad |\psi_\ast\rangle \qquad \},
\end{eqnarray*}
where $H[h_x(t)]$ is the Hamiltonian, see Sec.~\ref{sec:phases}, whose time dependence is defined through the magnetic filed $h_x(t)$ which the agent is constructing during the episode via online Q-Learning updates for specific single-particle and many-body examples.

Let us now briefly illustrate the protocol construction algorithm: for instance, if we start in the initial RL state $s_0\!=\!(t\!=\!0,h_x\!=\!-4)$, and take the action $a\!=\!\delta h_x \!=\! 8$, we go to the next RL state $s_1=(\delta t,+4)$. As a result of the interaction with the environment, the initial quantum state is evolved forward in time for one time step (from time $t_0=0$ to time $t_1=\delta t$) with the constant Hamiltonian $H[h_x=4]$: $|\psi(\delta t)\rangle = \mathrm{e}^{-iH[h_x=4]\delta t}|\psi_i\rangle$. After each step we compute the local reward according to Eq.~\eqref{eq:rewards}, and update the Q-function, even though the instantaneous reward at that step might be zero [the update will still be non-trivial in the later episodes, since information is propagated backwards from the end of the episode, see Eq.~\eqref{eq:bellman}]. This procedure is repeated until the end of the episode is reached at $t=T$. In general, one can imagine this partially-observable Markov decision process as a state-action-reward chain
\begin{equation*}
s_{0} \to a_{0}\to r_{0}\longrightarrow s_{1}\to a_{1}\to r_{1}\longrightarrow s_{2}\to \dots\longrightarrow  s_{N_T}.
\end{equation*}

The above paragraph explains how to choose actions according to some fixed policy $\pi(a|s)$ -- the probability of taking the action $a$ from the state $s$. Some RL algorithms, such as Policy Gradient directly optimize the policy. Instead, Watkins Q-Learning offers an alternative which allows to circumvent this. The central object in Q-Learning is the $Q(s,a)$ function which is given by the expected total return $R=\sum_{i=0}^{N_T}r_i$ at the end of each episode, starting from a fixed state $s$, taking the fixed action $a$, and acting optimally afterwards. Clearly, if we have the optimal $Q$-function $Q_\ast$, then the optimal policy is the deterministic policy $\pi_\ast(a|s)= 1,\ \mathrm{if}\ a=\argmax_{a'} Q(s,a')$, and $\pi_\ast(a|s)= 0$ for all other actions.

Hence, in Q-Learning one looks directly for the optimal Q-function. It satisfies the Bellman optimality equation, the solution of which cannot be obtained in a closed form for complex many-body systems~\footnote{Bellman's equation probably admits a closed solution for the single-qubit example from Sec.~\ref{sec:phases}.}. The underlying reason for this can be traced back to the non-integrability of the dynamical many-body system, as a result of which the solution of the Schr\"odinger equation cannot be written down as a closed-form expression even for a fixed protocol, and the situation is much more complicated when one starts optimizing over a family of protocols. The usual way of solving the Bellman equation numerically is Temporal Difference learning, which results in the following Q-Learning update rule~\cite{sutton_barto}
\begin{equation}
Q(s_i,a_i)\longleftarrow Q(s_i,a_i) + \alpha\left[ r_i \!+\! \max_a Q(s_{i\!+\!1},a) \!-\! Q(s_i,a_i) \right],
\label{eq:bellman}
\end{equation}
where the learning rate $\alpha\in(0,1)$. Whenever $\alpha\approx 1$, the convergence of the update rule~\eqref{eq:bellman} can be slowed down or even precluded, in cases where the Bellman error $\delta_t =  r_i + \max_a Q(s_{i+1},a) - Q(s_i,a_i)$ becomes significant. On the contrary, $\alpha\approx 0$ corresponds to very slow learning. Thus, the optimal value for the learning rate lies in between, and is determined empirically for the problem under consideration. 

To allow for the efficient implementation of piecewise-constant drives, i.e.~bang-bang protocols with a large number of bang modes, cf.~Ref.~\cite{supplementary}, we employ a linear function approximation to the $Q$-function, using equally-spaced tilings along the entire range of $h_x(t)\in[-4,4]$~\cite{sutton_barto}. The variational parameters of the linear approximator are found iteratively using Gradient Descent. This allows the RL agent to generalize, i.e.~gain information about the fidelity of not yet encountered protocols.

We iterate the algorithm for $2\times 10^4$ episodes. The exploration-exploitation dilemma~\cite{sutton_barto} requires a fair amount of exploration, in order to ensure that the agent visits large parts of the RL state space which prevents it from getting stuck in a local maximum of reward space from the beginning. Too much exploration, and the agent will not be able to learn. On the other hand, no exploration whatsoever guarantees that the agent will repeat deterministically a given policy, though it will be unclear whether there exists a better, yet unseen one. In the longer run, we cannot preclude the agent from ending up in a local maximum. In such cases, we run the algorithm multiple times starting from a random initial condition, and post-select the outcome. Hence, the RL solution is almost-optimal in the sense that its fidelity is close to the true global optimal fidelity. Unfortunately, the true optimal fidelity for nonintegrable many-body systems is unknown, and it is a definitive feature of glassy landscapes, see Sec.~\ref{sec:glassy}, that the true optimal is exponentially hard, and therefore also impractical, to find~\cite{day2018glassy}.

We also verified that RL does not depend on the initial condition chosen, provided the change is small. For instance, if one chooses different initial and target states which are both paramagnetic, then RL works with marginal drops in fidelity, which depend parametrically on the deviation from the initial and target states. If however, the target is, e.g.~paramagnetic and we choose an antiferromagnetic initial state [i.e.~the initial and target states are chosen in two different phases of matter], then we observe a drop in the found fidelity.

\begin{figure}[t!]
	\centering
	\begin{minipage}[b]{0.49\textwidth}
		\includegraphics[width=\textwidth]{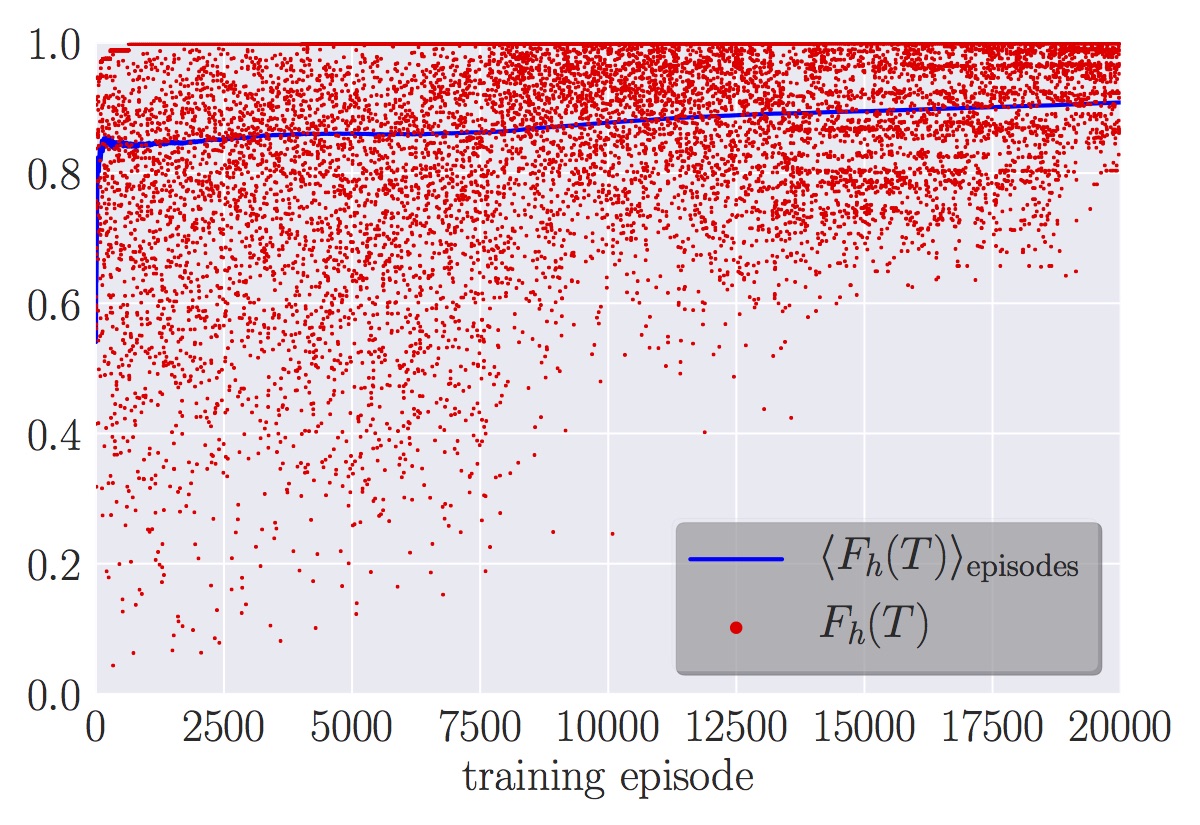}
	\end{minipage}
	\hfill
	\begin{minipage}[b]{0.49\textwidth}
		\includegraphics[width=\textwidth]{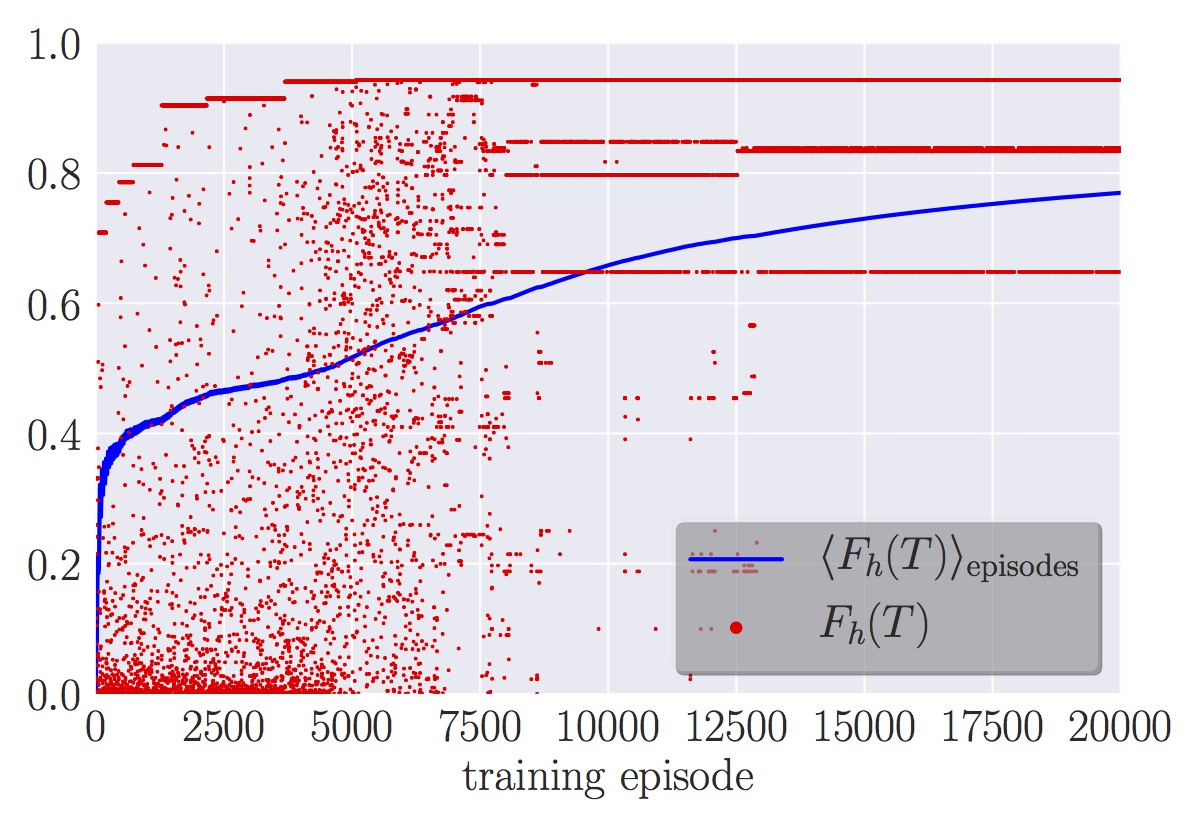}
	\end{minipage}
	\caption{\label{fig:RL_learning} Learning curves of the RL agent for the problems from Sec.~\ref{sec:phases} for $L=1$ at $T=2.4$ (up) [see \href{https://mgbukov.github.io/movies/RL_post/qubit.mp4}{Video 7}]  and $L=10$ at $T=3.0$ (down) [see \href{https://mgbukov.github.io/movies/RL_post/many_qubits.mp4}{Video 8}]. The red dots show the instantaneous reward (i.e.~fidelity) at every episode, while the blue line the cumulative episode-average. The ramp-up of the RL temperature $\beta_\mathrm{RL}$ gradually suppresses exploration over time which leads to a smoothly increasing average fidelity. The time step is $\delta t=0.05$.}
\end{figure} 

Due to the extremely large state space, we employ a replay schedule to ensure that our RL algorithm could learn from the high fidelity protocols it encountered. Our replay algorithm alternates between two different ways of training the RL agent which we call training stages: an ``exploratory'' training stage where the RL agent exploits the current $Q$-function to explore, and a ``replay'' training stage where we replay the best encountered protocol. This form of replay, to the best of our knowledge, has not been used previously. In the exploratory training stage, which lasts $40$ episodes, the agent takes actions according to a softmax probability distribution based on the instantaneous values of the $Q$-function. In other words, at each time step, the RL agent looks up the instantaneous values $Q(s,:)$ corresponding to all available actions, and computes a probability for each action: $P(a) \sim \exp(\beta_\mathrm{RL} Q(s,a))$. This exploration scheme results in random flips in the bangs of the protocol sequence, which is essentially a variation on the instantaneous RL best solution. Fig.~\ref{fig:RL_learning} shows that some of these variations lead to drastic reduction in fidelity, which we related to the glassy character of the correlated control phase, see Sec.~\ref{sec:glassy}. 

The amount of exploration is set by $\beta_\mathrm{RL}$, with $\beta_\mathrm{RL}=0$ corresponding to random actions and $\beta_\mathrm{RL}=\infty$ corresponding to always taking greedy actions with respect to the current estimate of the $Q$-function. Here we use an external `learning' temperature scale, the inverse of which, $\beta_\mathrm{RL}$, is linearly ramped down as the number of episodes progresses. In the replay training stage, which is also $40$ episodes long, we replay the best-encountered protocol up to the given episode. Through this procedure, when the next exploratory training stage begins again, the agent is biased to do variations on top of the best-encountered protocol, effectively improving it, until it reaches a reasonably good fidelity. 

Two learning curves of the RL agent are shown in Fig.~\ref{fig:RL_learning}. Notice the occurrence of suboptimal protocols even during later episodes due to the stochasticity of the exploration schedule. During every episode, the agent takes the best action (w.r.t.~its current knowledge/experience) with a finite probability, or else a random action is chosen. This prevents the algorithm from immediately getting stuck in a high-infidelity (i.e.~a bad) minimum. To guarantee convergence of the RL algorithm, the exploration probability is reduced as the number of episodes progresses (cf.~discussion above). This becomes manifest in Fig.~\ref{fig:RL_learning}, where after many episodes the deviations from the good protocols decrease. In the end, the agent learns the best-encountered protocol as a result of using the replay schedule which speeds up learning (as can be seen by the bad shots becoming rarer with increasing the number of episodes). We show only these learned protocols in Fig.~\ref{fig:phase_diag}b and Fig.~3 of the Supplemental Material~\cite{supplementary}.

\section{\label{sec:phases}Phases of Quantum Control}

\subsection{\label{subsec:singlequbit_diagram}Single Qubit Manipulation}

To benchmark the application of RL to physics problems, consider first a two-level system described by
\begin{equation}
	H[h_x(t)] = -S^z - h_x(t)S^x,
\label{eq:H_2LS}
\end{equation}
where $S^\alpha$, are the spin-$1/2$ operators. This Hamiltonian comprises both integrable many-body and non-interacting translational invariant systems, such as the transverse-field Ising model, graphene and topological insulators. The initial $|\psi_i\rangle$ and target $|\psi_\ast\rangle$ states are chosen as the ground states of~\eqref{eq:H_2LS} at $h_x=-2$ and $h_x=2$, respectively. We verified that the applicability of RL does not depend on this specific choice. Although there exists an analytical solution to solve for the optimal protocol in this case~\cite{hegerfeldt_13}, it does not generalize to non-integrable many-body systems. Thus, studying this problem using RL serves a two-fold purpose: (i) we benchmark the protocols obtained by the RL agent demonstrating that, even though RL is a completely model-free algorithm, it still finds the physically meaningful solutions by constructing a minimalistic effective model on-the-fly. The learning process is shown in \href{https://mgbukov.github.io/movies/RL_post/qubit.mp4}{Video 7}; (ii) We reveal an important novel perspective on the complexity of quantum state manipulation which, as we show below, generalizes to many-particle systems. While experimental set-ups studying single-qubit physics can readily apply multiple control fields (e.g also control fields in the $y$-direction) in order to test RL on a non-trivial problem with a known solution, we restrict the discussion to a single control parameter.

For fixed total protocol duration $T$, the infidelity $h_x(t)\mapsto I_h(T)=1-F_h(T)$ represents a ``potential landscape'', the global minimum of which corresponds to the optimal driving protocol. For bang-bang protocols, the problem of finding the optimal protocol becomes equivalent to finding the ground state configuration of a classical Ising model with complicated interactions~\cite{day2018glassy}. We map out the landscape of local infidelity minima $\{h_x^\alpha(t)\}_{\alpha=1}^{N_\mathrm{real}}$ using Stochastic Descent (SD), starting from random bang-bang protocol configurations~\cite{supplementary}. To study the correlations between the infidelity minima as a function of the total protocol duration $T$, we define the correlator $q(T)$, closely related to the Edwards-Anderson order parameter for the existence of spin glass order~\cite{castellani_05,hedges_09}, as
\begin{equation}
q(T)=\frac{1}{16N_T}\sum_{j=1}^{N_T}\overline{\{h_x(j\delta t)-\overline{h_x}(j\delta t)\}^2}, 
\label{eq:q_EA}
\end{equation} 
where $\overline{h_x}(t)=N_\mathrm{real}^{-1}\sum_{\alpha=1}^{N_\mathrm{real}}h_x^\alpha(t)$ is the sample-averaged protocol. If the minima $\{h_x^\alpha(t)\}_{\alpha=1}^{N_\mathrm{real}}$ are all uncorrelated, then $\overline{h_x}(t)\equiv 0$, and thus $q(T)=1$. On the other hand, if the infidelity landscape contains only one minimum, then $\overline{h_x}(t)\equiv h_x(t)$ and $q(T)=0$. The behaviour of $q(T)$, and the maximum fidelity $F_h(T)$ found using SD, together with a qualitative description of the corresponding infidelity landscapes are shown in Fig.~\ref{fig:phase_diag}.

The control problem for the constrained qubit exhibits three distinct control phases as a function of the protocol duration $T$. If $T$ is greater than the quantum speed limit $T_\mathrm{QSL}\approx 2.4$, one can construct infinitely many protocols which prepare the target state with unit fidelity, and the problem is in the \emph{controllable} phase $III$, c.f.~Fig.~\ref{fig:phase_diag}. The red line in Fig.~\ref{fig:phase_diag}b (iii) shows an optimal protocol of unit fidelity found by the agent, whose Bloch sphere representation can be seen in~\href{https://mgbukov.github.io/movies/RL_paper/Movie-3.mp4}{Video 3}. In this phase, there is a proliferation of \emph{exactly degenerate}, uncorrelated global infidelity minima, corresponding to protocols of unit fidelity, and the optimization task is easy.

At $T=T_\mathrm{QSL}$, the order parameter $q(T)$ exhibits a non-analyticity, and the system undergoes a continuous phase transition to a correlated  phase $II$. For times smaller than $T_\mathrm{QSL}$ but greater than $T_c$, the degenerate minima of the infidelity landscape recede to form a correlated  landscape with many \emph{non-degenerate} local minima, as reflected by the finite value of the order parameter $0<q(T)<1$. As a consequence of this correlated phase, there no longer exists a protocol to prepare the target state with unit fidelity, since it is physically impossible to reach the target state while obeying all constraints. The infidelity minimization problem is non-convex, and determining the best achievable (i.e.~optimal) fidelity [a.k.a.~the global minimum] becomes difficult. Figure~\ref{fig:phase_diag}b (ii)shows the best bang-bang protocol found by our computer agent (see~\href{https://mgbukov.github.io/movies/RL_paper/Movie-2.mp4}{Video 2} and~\cite{supplementary} for protocols with quasi-continuous actions). This protocol has a remarkable feature:  without any prior knowledge about the intermediate quantum state nor its Bloch sphere representation, the model-free RL agent discovers that it is advantageous to first bring the state to the equator -- which is a geodesic -- and then effectively turns off the control field $h_x(t)$, to enable the fastest possible precession about the $z$-axis \footnote{Notice that the agent does not have control over the $z$ field.}. After staying on the equator for as long as optimal, the agent rotates as fast as it can to bring the state as close as possible to the target, thus optimizing the final fidelity for the available protocol duration. 

Decreasing the total protocol duration $T$ further, we find a second critical time $T_c\approx 0.6$. For $T<T_c$, $q(T)\equiv 0$ and the problem has a unique solution, suggesting that the infidelity landscape is convex. This \emph{overconstrained phase} is labelled $I$ in the phase diagram (Fig.~\ref{fig:phase_diag}a). For $T<T_c$, there exists a unique optimal protocol, even though the achievable fidelity can be quite limited, see Fig.~\ref{fig:phase_diag}b (i) and \href{https://mgbukov.github.io/movies/RL_paper/Movie-1.mp4}{Video 1}. Since the state precession speed towards the equator depends on the maximum possible allowed field strength $h_x$, it follows that $T_c\to 0$ for $|h_x|\to\infty$. 

\begin{figure}[t!]
	\includegraphics[width=0.49\textwidth]{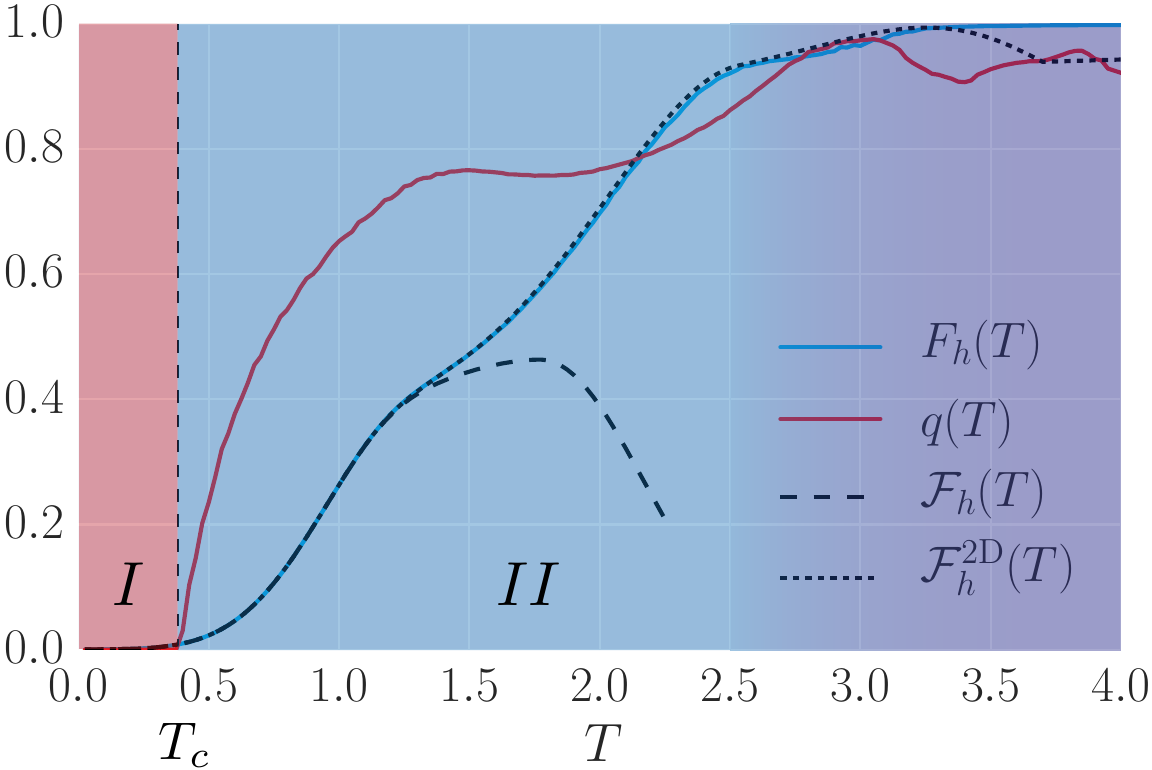}
	\caption{\label{fig:MB_phasediag} Phase diagram of the many-body quantum state manipulation problem. The order parameter (red) shows a kink at the critical time $T_c\approx 0.4$ when a phase transition occurs from an overconstrained phase ($I$) to a glassy phase ($II$). The best fidelity $F_h(T)$ (blue) obtained using SD is compared to the variational fidelity $\mathcal{F}_h(T)$ (dashed) and the 2D-variational fidelity $\mathcal{F}^\mathrm{2D}_h(T)$ (dotted)~\cite{supplementary}.}
\end{figure}

\emph{Relation to Counter-Diabatic and Fast-Forward Driving.---}Promising analytical approaches to state manipulation have recently been proposed, known as Shortcuts to Adiabaticity~\cite{demirplak_03,demirplak_05,demirplak_08,berry_09,delcampo_13,deffner_14,kolodrubetz_16,sels_16,theisen_17,petiziol2018fast}. They include ideas such as (i) fast-forward (FF) driving, which comprises protocol that excite the system during the evolution at the expense of gaining speed, before taking away all excitations and reaching the target state with unit probability, and (ii) counter-diabatic (CD) driving, which ensures transitionless dynamics by turning on additional control fields. In general any FF protocol is related to a corresponding CD protocol. While for complex many-body systems, it is not possible to construct the mapping between FF and CD in general, the simplicity of the single-qubit setup~\eqref{eq:H_2LS} allows to use CD driving to find a FF protocol~\cite{bukov_GSL}. For an unbounded control field $h_x(t)$, the FF protocol at the quantum speed limit has three parts which can be understood intuitively on the Bloch sphere: (i) an instantaneous delta-function kick to bring the state to the equator, (ii) an intermediate stage where the control field is off, $h_x(t)\equiv 0$, which allows the state to precess along the equator, and (iii) a complementary delta kick to bring the state from the equator straight to the target~\cite{bukov_GSL}. Whenever the control field is bounded, $|h_x|\leq 4$, these delta kicks are broadened and take extra time, thus increasing $T_\mathrm{QSL}$. If the RL algorithm finds a unit-fidelity protocol, it is by definition a FF one. Comparing FF driving to the protocol found by our RL agent [cf.~Fig.~\ref{fig:phase_diag}b, see also paragraphs above], we find indeed a remarkable similarity between the RL and FF protocols.

\subsection{\label{subsec:manyqubit_diagram}Many Coupled Qubits}

The above results raise the natural question of how much more difficult state manipulation is in more complex quantum models. To this end, consider a closed chain of $L$ coupled qubits, which can be experimentally realized with superconducting qubits~\cite{barends_16}, cold atoms~\cite{simon_11} and trapped ions~\cite{jurcevic_14}:
\begin{equation}
H[h_x(t)] = -\sum_{j=1}^L \left(S^z_{j+1}S^z_j + gS^z_j +h_x(t)S^x_j\right).
\label{eq:H_MB}
\end{equation}
We set $g\!=\!1$ to avoid the anti-ferromagnet to paramagnet phase transition, and choose the paramagnetic ground states of Eq.~\eqref{eq:H_MB} at fields $h_x=-2$ and $h_x=2$ for the initial and target state, respectively. We verified that the conclusions we draw below do not depend on the choice of initial and target states, provided they both belong to the paramagnetic phase. The details of the control field $h_x(t)$ are the same as in the single qubit case, and we use the \emph{many-body} fidelity both the reward and the measure of performance. In this paper, we focus on $L>2$. The two-qubit optimization problem was shown to exhibit an additional symmetry-broken correlated phase, see Ref.~\cite{bukov_17symmbreak}.

Figure~\ref{fig:MB_phasediag} shows the phase diagram of the coupled qubits model. First, notice that while the overconstrained-to-glassy critical point $T_c$ survives, the quantum speed limit critical point $T_\mathrm{QSL}$ is (if existent at all) outside the short protocol-time range of interest. Thus, the glassy phase extends over to long and probably infinite protocol durations, which offers an alternative explanation for the difficulty of preparing many-body states with high fidelity. The glassy properties of this phase are analyzed extensively in Ref.~\cite{day2018glassy}. Second, observe that, even though unit fidelity is no longer achievable, there exist nearly optimal protocols with extremely high many-body fidelity \footnote{We expect the \emph{many-body} overlap between two states to be exponentially small in the system size $L$.} at short protocol durations. This fact is striking because the Hilbert space of our system grows \emph{exponentially} with $L$ and we are using only one control field to manipulate exponentially many degrees of freedom in a short time. Nonetheless, it has been demonstrated that two states very close to each other in, or of equal, fidelity can possess sufficiently different physical properties or be very far in terms of physical resources~\cite{benedetti_13,bina_14,mandarino_16,bukov_17symmbreak}. Hence, one should be cautious when using the fidelity as a measure for preparing many-body states and exploring other possible reward functions for training RL agents is an interesting avenue for future research. 

Another remarkable characteristic of the optimal solution is that for the system sizes $L\geq 6$ both $q(T)$ and $-L^{-1}\log F_h(T)$ converge to their thermodynamic limit values with no visible finite-size corrections~\cite{supplementary}. This is likely related to the Lieb-Robinson bound for information propagation which suggests that information should spread over approximately $JT=4$ sites for the longest protocol durations considered. 

\section{\label{sec:variational}Variational Theory for Nearly-Optimal Protocols}

An additional feature of the optimal bang-bang solution found by the agent is that the entanglement entropy of the half system generated during the evolution always remains small, satisfying an area law~\cite{supplementary}. This implies that the system likely follows the ground state of some local, yet a-priori unknown effective Hamiltonian~\cite{vidmar_15}. This emergent behavior motivated us to use the best protocols found by ML to construct simple variational protocols consisting of just a few bangs. Let us now demonstrate how this works by giving specific examples which, to our surprise, capture the essence of the phase diagram of quantum control both qualitatively and quantitatively. 

\subsection{\label{subsec:var_single_qubit}Single Qubit}

By carefully studying the optimal driving protocols the RL agent finds in the case of the single qubit, we find a few important features. Focussing for the moment on bang-bang protocols, in the overconstrained and correlated phases [cf.~Fig.~\ref{fig:phase_diag}b and \href{https://mgbukov.github.io/RL_movies/}{Videos 1--3}], we recognize an interesting pattern: for $T<T_c$, as we explained in Sec.~\ref{sec:phases}, there is only one minimum in the infidelity landscape, which dictates a particularly simple form for the bang-bang protocol -- a single jump at half the total protocol duration $T/2$. On the other hand, for $T_c\leq T\leq T_\mathrm{QSL}$, there appears a sequence of multiple bangs around $T/2$, which grows with increasing the protocol duration $T$. By looking at the Bloch sphere representation, see~\href{https://mgbukov.github.io/RL_movies/}{Videos 1--3}, we identify this as an attempt to turn off the $h_x$-field, once the state has been rotated to the equator. This trick allows for the instantaneous state to be moved in the direction of the target state in the shortest possible distance [i.e.~along a geodesic].

\begin{figure*}[t!]	
	\includegraphics[width=0.9\textwidth]{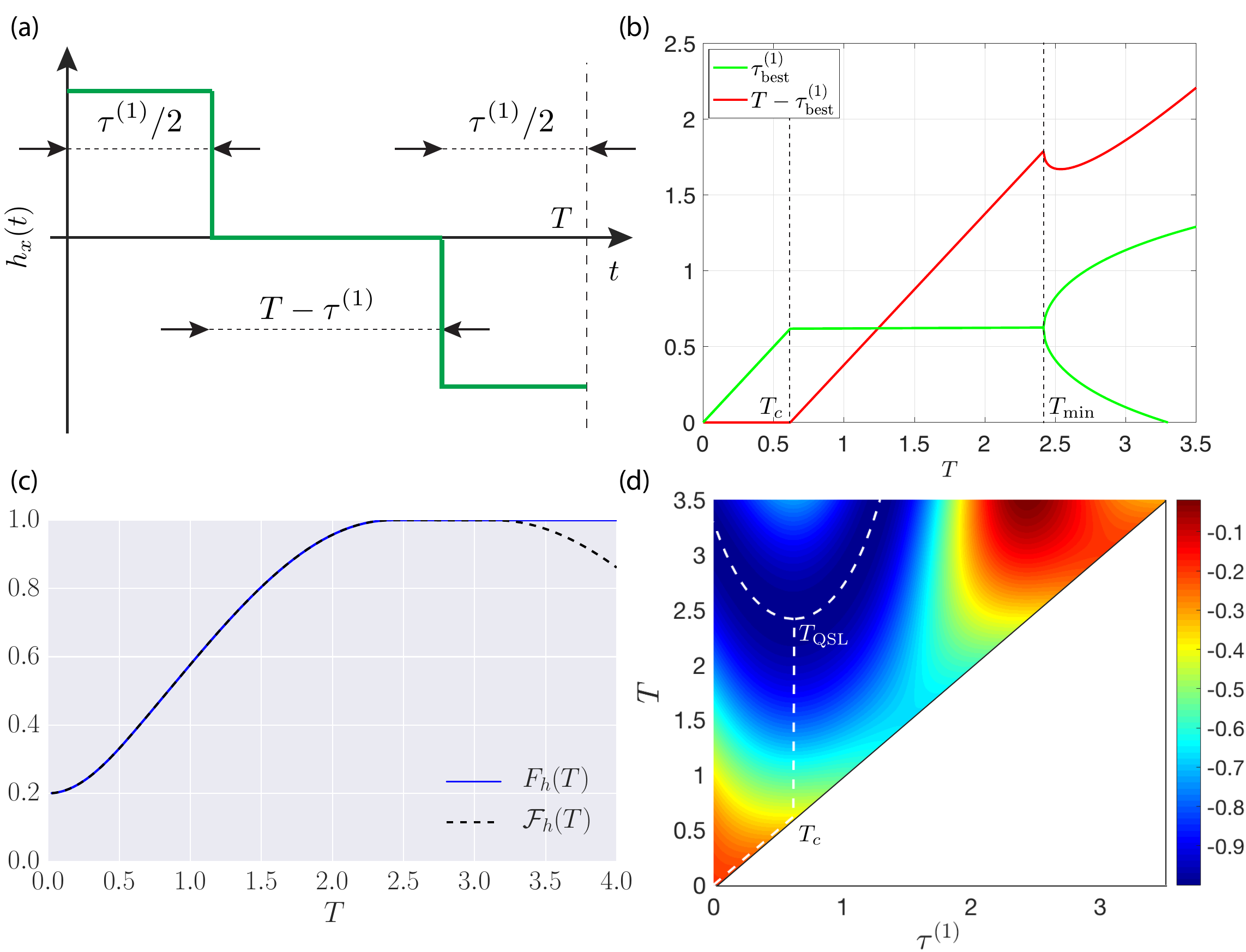}
	\caption{\label{fig:2LS_theory} (a) Three-pulse variational protocol which allows to capture the optimal protocol found by the computer in the overconstrained and the glassy phases of the single qubit problem. (b) $\tau^{(1)}_\mathrm{best}$ (green), with the non-analytic points of the curve marked by dashed vertical lines corresponding to $T_c\approx 0.618$ and $T_\mathrm{QSL}\approx 2.415$. (c) Best fidelity obtained using SD (solid blue) and the variational ansatz (dashed black). (d) The variational infidelity landscape with the minimum for each $T$-slice designated by the dashed line which shows the robustness of the variational ansatz against small perturbations.}
\end{figure*}

Hence, it is suggestive to try out a three-pulse protocol as an ansatz for the optimal solution, see Fig.~\ref{fig:2LS_theory}a: the first (positive) pulse of duration $\tau^{(1)}/2$ brings the state to the equator. Then the $h_x$-field is turned off for a time $\tilde\tau^{(1)}=T-\tau^{(1)}$, after which a negative pulse directs the state off the equator towards the target state. Since the initial value problem is time-reversal symmetric for our choice if initial and target states, the duration of the third pulse must be the same as that of the first one. We thus arrive at a variational protocol, parametrised by $\tau^{(1)}$, see Fig.~\ref{fig:2LS_theory}a.

The optimal fidelity is thus approximated by the variational fidelity $\mathcal{F}_h(\tau^{(1)},T-\tau^{(1)})$ for the trial protocol [Fig.~\ref{fig:2LS_theory}a], and can be evaluated analytically in a straightforward manner:
\begin{widetext}
\begin{eqnarray}
\mathcal{F}_h(\tau^{(1)},T-\tau^{(1)}) &=& |\langle\psi_\ast|\mathrm e^{-i\frac{\tau^{(1)}}{2}H[-h_\mathrm{max}]} \mathrm e^{-i(T-\tau^{(1)}) H[0]}\mathrm e^{-i\frac{\tau^{(1)}}{2}H[h_\mathrm{max}]}|\psi_i\rangle|^2,\nonumber\\
H[h_x] &=& -S^z - h_xS^x.
\end{eqnarray} 
\end{widetext}
However, since the exact expression is rather cumbersome, we choose not to show it explicitly. Optimizing the variational fidelity at a fixed protocol duration $T$, we solve the corresponding transcendental equation to find the extremal value $\tau^{(1)}_\mathrm{best}$, and the corresponding optimal variational fidelity $\mathcal{F}_h(T)$, shown in Fig.~\ref{fig:2LS_theory}b-c. For times $T\leq T_c$, we find $\tau^{(1)}=T$ which corresponds to $\tilde\tau^{(1)}=0$, i.e.~a single bang in the optimal protocol. The overconstrained-to-correlated phase transition at $T_c$ is marked by a non-analyticity at $\tau^{(1)}_\mathrm{best}(T_c) = T_c \approx 0.618$. This is precisely the minimal time the agent can take, to bring the state to the equator of the Bloch sphere, and it depends on the value of the maximum magnetic field allowed [here $h_\mathrm{max}=4$]. Figure~\ref{fig:2LS_theory}d shows that, in the overconstrained phase, the fidelity is optimised at the boundary of the variational domain, although $\mathcal{F}_h(\tau^{(1)},T-\tau^{(1)})$ is a highly nonlinear function of $\tau^{(1)}$ and $T$.

For $T_c\leq T\leq T_\mathrm{QSL}$, the time $\tau^{(1)}$ is kept fixed [the equator being the only geodesic for a rotation along the $\hat z$-axis of the Bloch sphere], while the second pulse time $\tilde\tau^{(1)}$ grows linearly, until the minimum time $T_\mathrm{QSL}\approx 2.415$ is eventually reached. The minimum time is characterised by a bifurcation in our effective variational theory, as the corresponding variational infidelity landscape develops two minima, see Fig.~\ref{fig:2LS_theory}b,d. Past that protocol duration, our simplified ansatz is no longer valid, and the system is in the controllable phase. Furthermore, a sophisticated analytical argument based on optimal control theory can give exact expressions for $T_c$ and $T_\mathrm{QSL}$~\cite{hegerfeldt_13}, in precise agreement with the values we obtained. The Bloch sphere representation of the variational protocols in Fig.~\ref{fig:phase_diag}b (dashed blue lines) for the single qubit are shown in~\href{https://mgbukov.github.io/RL_movies/}{Videos 4-6}.

\begin{figure*}[t!]	
	\includegraphics[width=1.0\textwidth]{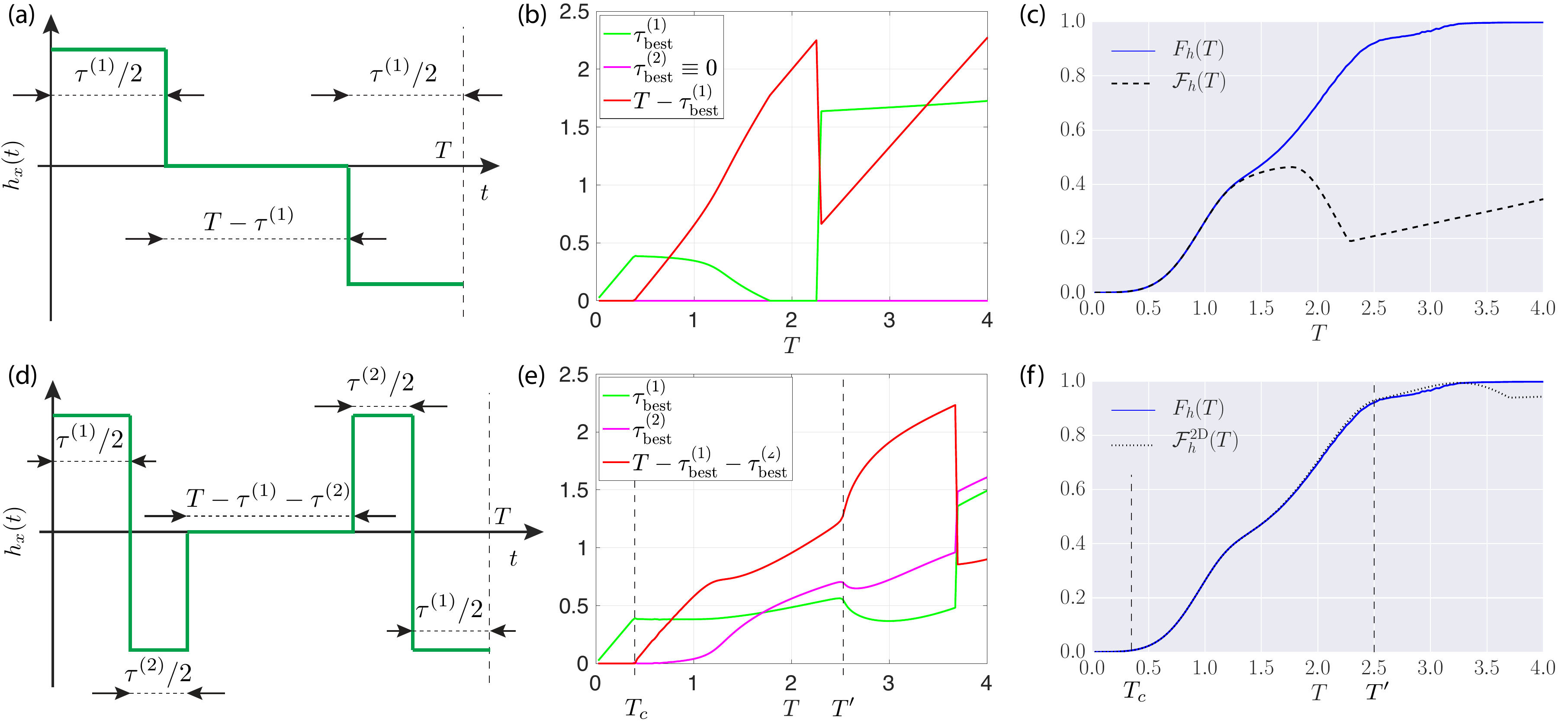}
	\caption{\label{fig:MB_theory} 	
		(a) Three-pulse variational protocol which allows to capture the optimal protocol found by the computer in the overconstrained phase but fails the glassy phase of the nonintegrable many-body problem. This ansatz captures the non-analytic point at $T_c\approx 0.4$ but fails in the glassy phase. (b) The pulse durations $\tau^{(1)}_\mathrm{best}$ (green) and $\tau^{(2)}_\mathrm{best}$ (magenta), for highest fidelity variational protocol of length $T$ of the type shown in (a). The fidelity of the variational protocols exhibit a physical non-analyticity at $T_c\approx 0.4$ and unphysical kinks outside the validity of the ansatz. (c) 1D maximal variational fidelity (dashed back) compared to the best numerical protocol (solid blue).
		(d) Five-pulse variational protocol which allows to capture the optimal protocol found by the computer in the overconstrained phase and parts of the glassy phase of the nonintegrable many-body problem. (e) The pulse durations $\tau^{(1)}_\mathrm{best}$ (green) and $\tau^{(2)}_\mathrm{best}$ (magenta) for the best variational protocol of length $T$ of the type shown in (d). These variational protocols exhibit physical non-analyticities at $T_c\approx 0.4$ and $T'\approx 2.5$ (vertical dashed lines) (f) 2D maximal variational fidelity (dashed-dotted back) compared to the best numerical protocol (solid blue).}
\end{figure*}

To summarize, for the single qubit example, the variational fidelity $\mathcal{F}_h(T)$ agrees nearly perfectly with the optimal fidelity $F_h(T)$ obtained using SD and Optimal Control, cf.~Fig.~\ref{fig:phase_diag}a. We further demonstrate that our variational theory fully captures the physics of the two critical points $T_c$ and $T_\mathrm{QSL}$~\cite{supplementary}. Interestingly, the variational solution for the single qubit problem coincides with the global minimum of the infidelity landscape all the way up to the quantum speed limit~\cite{supplementary}.

\subsection{\label{subsec:var_many_qubits}Many Coupled Qubits}

Let us also discuss the variational theory for the many-body system. Consider first the same one-parameter variational ansatz from Sec.~\ref{subsec:var_single_qubit}, see Fig.~\ref{fig:MB_theory}a. Since the variational family is one-dimensional, we shall refer to this ansatz as the 1D variational theory. The dashed black line in Fig.~\ref{fig:MB_theory}c shows the corresponding 1D variational fidelity. We see that, once again, this ansatz captures correctly the critical point $T_c$ separating the overconstrained and the glassy phases. Nevertheless, a comparison with the optimal fidelity [see Fig.~\ref{fig:MB_theory}c] reveals that this variational ansatz breaks down in the glassy phase, although it rapidly converges to the optimal fidelity with decreasing $T$. Looking at Fig.~\ref{fig:MB_theory}b, we note that the value $\tau^{(1)}_\mathrm{best}$, which maximizes the variational fidelity, exhibits a few kinks. However, only the kink at $T=T_c$ captures a physical transition of the original control problem, while the others appear as artefacts of the simplified variational theory, as can be seen by the regions of agreement between the optimal and variational fidelities.

Inspired by the structure of the protocols found by our RL agent once again, see \href{https://mgbukov.github.io/movies/RL_post/many_qubits.mp4}{Video 8}, we now extend the qubit variational protocol, as shown in Fig.~\ref{fig:MB_theory}d. In particular, we add two more pulses to the protocol, retaining its symmetry structure: $h_x(t)= -h_x(T-t)$, whose length is parametrised by a second, independent variational parameter $\tau^{(2)}/2$. Thus, the pulse length where the field is set to vanish, is now given by $\tilde{\tau}=T-\tau^{(1)}-\tau^{(2)}$. These pulses are reminiscent of spin-echo protocols, and appear to be important for entangling and disentangling the state during the evolution. Notice that this extended variational ansatz includes by definition the simpler ansatz from the single qubit problem discussed above, by setting $\tau^{(2)}=0$. 

Let us now turn on the second variational parameter $\tau^{(2)}$, and consider the full two-dimensional variational problem:
\begin{widetext} 
\begin{eqnarray}
\mathcal{F}^\mathrm{2D}_h(\tau^{(1)},\tau^{(2)},T-\tau^{(1)}-\tau^{(2)}) &=&
|\langle\psi_\ast|\mathrm e^{-i\frac{\tau^{(1)}}{2}H[-h_\mathrm{max}]} 
\mathrm e^{-i\frac{\tau^{(2)}}{2}H[h_\mathrm{max}]}
\mathrm e^{-i(T-\tau^{(1)}) H[0]}
\mathrm e^{-i\frac{\tau^{(2)}}{2}H[-h_\mathrm{max}]}
\mathrm e^{-i\frac{\tau^{(1)}}{2}H[h_\mathrm{max}] }|\psi_i\rangle|^2,\nonumber\\
H[h_x] &=& -\sum_{j=1}^L \left(S^z_{j+1}S^z_j + gS^z_j +h_xS^x_j\right).
\end{eqnarray}
\end{widetext}
For the maximum-fidelity variational protocol, we show the best variational fidelity $\mathcal{F}^\mathrm{2D}_h$ [Fig.~\ref{fig:MB_theory}f] and the corresponding values of $\tau^{(1)}_\mathrm{best}$ and $\tau^{(2)}_\mathrm{best}$ [Fig.~\ref{fig:MB_theory}e]. There are two important points here: (i) Fig.~\ref{fig:MB_theory}f shows that the 2D variational fidelity seemingly reproduces the optimal fidelity on a much longer scale compared to the 1D variational ansatz, i.e.~for all protocol durations $T\lesssim 3.3$. (ii) the 2D variational ansatz reduces to the 1D one in the overconstrained phase $T\leq T_c$. In particular, both pulse lengths $\tau^{(1)}_\mathrm{best}$ and $\tau^{(2)}_\mathrm{best}$ exhibit a non-analyticity at $T=T_c$, but also at $T'\approx 2.5$. Interestingly, the 2D variational ansatz captures the optimal fidelity on both sides of $T'$ which suggests that there is likely yet another transition within the glassy phase, hence the different shading in the many-body phase diagram [Fig.~\ref{fig:MB_phasediag}]. Similar to the 1D variational problem, here we also find artefact transitions [non-analytic behavior in $\tau^{(i)}_\mathrm{max}$ outside of the validity of the variational approximation]. 

In summary, in the many-body case, the same one-parameter variational ansatz only describes the behaviour in the overconstrained phase, cf.~Fig.~\ref{fig:MB_phasediag} (dashed line), up to and including the critical point $T_c$, but fails for $T>T_c$. Nevertheless, a slightly modified, two-parameter variational ansatz, motivated again by the solutions found by the ML agent (see \href{https://mgbukov.github.io/movies/RL_post/many_qubits.mp4}{Video 8}), appears to be fully sufficient to capture the essential features of the optimal protocol much deeper into the glassy phase, as shown by the $\mathcal{F}^\mathrm{2D}_h(T)$ curve in Fig.~\ref{fig:MB_phasediag}. This many-body variational theory features an additional pulse, reminiscent of spin-echo, which appears to control and suppress the generation of entanglement entropy during the drive~\cite{supplementary}. Indeed, while the two-parameter ansatz is strictly better than the single-parameter protocol for all $T>T_c$, the difference between the two grows slowly as a function of time. It is only at a later time, $T\approx 1.3$, that the effect of the second pulse really kicks in, and we observe the largest entanglement in the system for the optimal protocol.

Using RL, we identified nearly-optimal control protocols~\cite{gingrich_16} which can be parametrized by a few d.o.f. Such simple protocols have been proven to exist in weakly-entangled one-dimensional spin chains~\cite{lloyd_14}. However, the proof of the existence does not imply that these d.o.f.~are easy to identify. Initially, the RL agent is completely ignorant about the problem and explores many different protocols, while it tries to learn the relevant features. In contrast, optimal control methods, such as CRAB~\cite{caneva_11}, usually have a much more rigid framework, where the d.o.f.~of the method are fixed from the beginning. This can limit the performance of those methods below the quantum speed limit~\cite{sorensen_16,supplementary}. 

One might wonder how the nearly-optimal protocols found using RL and SD correlate with the best variational protocols. For the problem under consideration, averaging parts of the set of bang-bang protocols, which contains randomly generated local minima of the infidelity landscape, $\{h_x^\alpha(t)\}_{\alpha=1}^{N_\mathrm{real}}$, (see insets in Fig.~\ref{fig:SD_traces}) results in protocols which resemble the continuous ones we found using GRAPE. The variational solutions are indeed close to these averaged solutions, although they are not exactly the same, since the variational protocols are constrained to take on three discrete values (positive, zero and negative), while the averaged protocols can take on any values in the interval $[-4, 4]$. The RL agent cannot find these variational solutions because we have limited the actions space to having $h_x$ take the minimum or maximum allowable value and there is no way to take an action where $h_x=0$. 

We also showed how, by carefully studying the driving protocols found by the RL agent, one can obtain ideas for effective theories which capture the essence of the underlying physics. This approach is similar to using an effective $\phi^4$-theory to describe the physics of the Ising phase transition. The key difference is that the present problem is out-of-equilibrium, where no general theory of statistical mechanics exists so far. We hope that, an underlying pattern between such effective theories can be revealed with time, which might help shape the guiding principles of a theory of statistical physics away from equilibrium.

\section{\label{sec:glassy}Glassy Behaviour}

\begin{figure}[t!]
	\includegraphics[width=0.49\textwidth]{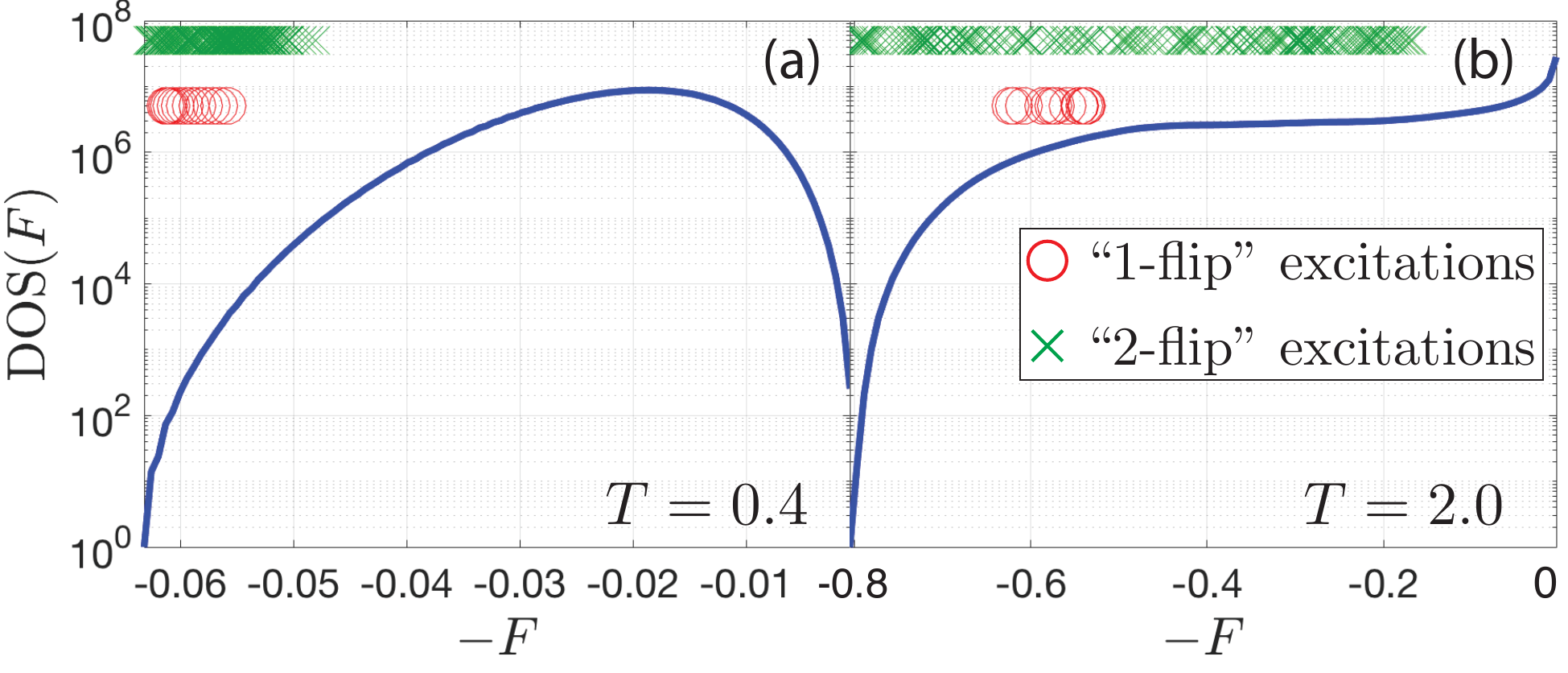}
	\caption{\label{fig:glass_main} Density of states (protocols) in the overconstrained phase at $T=0.4$ (a) and the glassy phase at $T=2.0$ (b) as a function of the fidelity $F$. The red circles and the green crosses show the fidelity of the ``1-spin'' flip and ``2-spin'' flip excitation protocols above the absolute ground state (i.e.~the optimal protocol). The system size is $L=6$ and each protocol has $N_T=28$ bangs. }
\end{figure}

It is quite surprising that the dynamics of a non-integrable many-body quantum system, associated with the optimal protocol, is so efficiently captured by such a simple, two-parameter variational protocol, even in the regimes where there is no obvious small parameter and where spin-spin interactions play a significant role. Upon closer comparison of the variational and the optimal fidelities, one can find regions in the glassy phase where the simple variational protocol outperforms the numerical `best' fidelity, cf.~Fig.~\ref{fig:MB_phasediag}.

To better understand this behavior, we choose a grid of $N_T=28$ equally-spaced time steps, and compute all $2^{28}$ bang-bang protocols and their fidelities. The corresponding density of states (DOS) in fidelity space is shown in Fig.~\ref{fig:glass_main} for two choices of $T$ in the overconstrained and glassy phase. This allows us to unambiguously determine the ground state of the infidelity landscape (i.e.~the optimal protocol). Starting from this ground state, we then construct all excitations generated by local in time flips of the bangs of the optimal protocol. The fidelity of the ``1-flip" excitations is shown using red circles in Fig.~\ref{fig:glass_main}. Notice how, in the glassy phase, these $28$ excitations have relatively low fidelities compared to the ground state, and are surrounded by $\sim 10^6$ other states. This has profound consequences: as we are `cooling' down in the glassy phase, searching for the optimal protocol and coming from a state high up in the infidelity landscape, if we miss one of the $28$ elementary excitations, it becomes virtually impossible to reach the global ground state and the situation becomes much worse if we increase the number of steps $N_T$. On the contrary, in the overconstrained phase, the smaller value of the DOS at the ``1-flip'' excitation ($\sim 10^2$) makes it easier to reach the ground state. 

The green crosses in Fig.~\ref{fig:glass_main} show the fidelity of the ``2-flip" excitations. By the above argument, a ``2-flip" algorithm would not see the phase as a glass for $T\lesssim 2.5$, yet it does so for $T\gtrsim 2.5$, marked by the different shading in Fig.~\ref{fig:MB_phasediag}. Correlated with this observation, we find a signature of a transition also in the improved two-parameter variational theory in the glassy phase [see Sec.~\ref{subsec:var_many_qubits} and kinks at $T'$ in Fig.~\ref{fig:MB_theory}e]. In general, we expect the glassy phase to exhibit a series of phase transitions, reminiscent of the random $k$-SAT problems~\cite{mezard_02,morampudi_17}. The glassy nature of this correlated phase has been studied in detail in Ref.\cite{day2018glassy}, by mapping this optimal control problem to an effective classical spin energy function which governs the control phase transitions.

\begin{figure}[t!]
	\centering
	\includegraphics[scale=0.7]{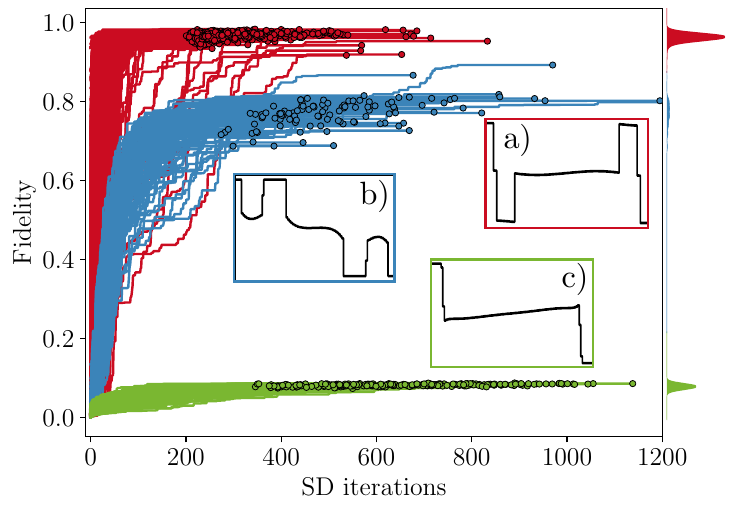}
	\caption{Fidelity traces of Stochastic Descent (SD) for $T=3.2, L=6$ and $N_T=200$ as a function of the number of iterations of the algorithm for $10^3$ random initial conditions. The traces are characterized by three main attractors marked by the different colors. The termination of each SD run is indicated by a colored circle. The relative population of the different attractors is shown as a density profile on the right-hand side. Inset (a)-(b)-(c): averaged profile of the protocols obtained for the red, blue and green attractor respectively.
		\label{fig:SD_traces}}
\end{figure}

In contrast to the single-qubit system, there are also multiple attractors present in the glassy phase of the many-body system [see Fig.~\ref{fig:SD_traces}]. Each attractor has a typical representative protocol [Fig.~\ref{fig:SD_traces} insets]. Even though intra-attractor protocols share the same averaged profile, they can nevertheless have a small mutual overlap -- comparable to the overlap of inter-attractor protocols. This indicates that in order to move in between protocols within an attractor, highly non-local moves are necessary. For this reason, GRAPE~\cite{grape_05}, an algorithm which performs global updates on the protocol by computing exact gradients in the control landscape, also performs very well on our optimisation problem. Similar to SD, in the glassy phase GRAPE cannot escape local minima in the infidelity landscape and, therefore, the same three attractors are found with comparable relative populations to SD, but intra-attractor fluctuations are significantly suppressed due to GRAPE's non-local character.
 
\section{\label{sec:outro}Outlook \& Discussion}

In this work we demonstrated the usefulness of Q-Learning to manipulate single-particle and many-body quantum systems. Q-Learning is only one of many Reinforcement Learning algorithms, including SARSA, Policy Gradient and Actor Critic methods, just to name a few. In the Supplemental Material we showed that Q-Learning's performance is comparable to many of the leading Optimal Control algorithms~\cite{supplementary}. It will be interesting and desirable to compare different RL algorithms among themselves on physical quantum systems. An exciting future direction is to investigate which advantages Deep Learning offers in the context of quantum control, and there exist recent studies exploring Deep RL in a physics~\cite{foesel_18,august_18,niu2018universal}.

Looking forward to controlling non-integrable many-body systems, an important question arises as to how the computational cost of Q-Learning scales with the system size $L$. As we explained in Sec.~\ref{sec:RL}, the Q-Learning algorithm can be decomposed into a `learning' part, and an `interaction with the environment' part where all physics/dynamics happens. The learning part does not know about the state of the quantum system -- it only keeps track of the value of the magnetic field at a given time $(t,h_x(t))$. As a result of this choice, for a single global drive the `learning part' of the algorithm is independent of the system size $L$ since it depends  only on a single scalar reward - the fidelity of the final state. The RL algorithm is instead computationally limited by the size of the action and state spaces. As currently implemented, this means that the RL algorithm is limited to finding short protocols (since the state space scales exponentially with the number of bangs). However, it may be possible to circumvent this bottleneck by using Deep RL which uses neural networks to represent the Q-function.

One place where the system size implicitly enters the computational costs of the RL protocol is through the number of episodes needed to train the RL algorithm. At every time step, one solves Schr\"odinger's equation to simulate the dynamics. The solver's scaling with $L$ depends on how the time evolution is implemented: in spin-$1/2$ systems, for exact diagonalization [used here] the computational cost scales exponentially $2^{2L}$, while a more sophisticated Krylov method alleviates this somewhat to $L^2 2^L$~\cite{saad1992analysis}, and matrix product states only scale as $L^2$ [in the worst case]~\cite{garcia2006time}. Therefore, combining RL with existing approximate techniques to evolve quantum states can lead to a significant reduction of CPU time, provided applying these techniques is justified by the underlying physics.

The present work demonstrates the suitability of RL for manipulating/controlling quantum systems. Yet, it does not explore how one can improve the Q-Learning algorithm, and adjust it to the specific needs of quantum control. Let us briefly list a few possible directions that the interested reader may want to keep in mind: (i) Alternative definitions of the RL state space, see Ref.~\cite{sutton_barto}, may prove advantageous, depending on the need of the problem setup, since this defines the agent's knowledge about the physical system. For instance, if the RL agent is to be coupled to an experiment, one cannot use the wavefunction for this purpose, whereas wavefunctions may be accessible in numerical simulations. We find that the choice of RL state space influences the learning capabilities of the RL agent.(ii) Another way to increase performance is to add more controls. This only increases the possibility to reach a higher fidelity, but it comes at a cost of a potential slow-down, due to a higher computational demand to explore the increased RL state space. (iii) In addition, choosing a suitable family of protocols and how to parametrize it, may also lead to increased performance in RL. We used bang-bang protocols because of their computational simplicity, yet the needs of a given problem may justify another choice: the experimental realization of bang-bang protocols is limited by the resolution with which a pulse can be stabilized, which is set by the experimental apparatus. Alternatively, the RL setup can be formulated to control the size of some generalized Fourier coefficients, an idea underlying the CRAB algorithm. (iv) On the algorithmic side, one can also optimize the exploration and replay schedules which control the learning efficiency with increasing the number of training episodes, and influence the RL agent's learning speed.

Reinforcement Learning algorithms are versatile enough and can be suitably combined with existing ones. For instance, applying RL to complex problems with glassy landscapes is likely to benefit from a pre-training stage. Such a beneficial behaviour has already been observed in the context of deep RL~\cite{mnih_15_ATARI,silver_16_ALPHAGO}. For the purpose of pre-training, in certain cases it may be advantageous to combine RL with existing derivative-based optimal control methods, such as GRAPE and CRAB, or even exhaustive search, so that one starts the optimization from a reasonable `educated guess'. In the recent years, it was shown that derivative-based and feedback-loop control methods can be efficiently combined to boost performance~\cite{egger2014adaptive}. Vice-versa, RL's exploration schedule defined on a suitable abstract RL-state space, may prove a useful addition to improve on already existing algorithms.  

Using RL, we revealed the existence of control phase transitions, and showed their universality in the sense that they also affect the behaviour of state-of-the-art optimal control methods. The appearance of a glassy phase, which dominates the many-body physics, in the space of protocols of the quantum state manipulation problem, could have far-reaching consequences for efficiently manipulating systems in condensed matter experiments. Quantum computing relies heavily on our ability to prepare states with high fidelity, yet finding high efficiency state manipulation routines remains a difficult problem. Highly controllable quantum emulators, such as ultracold atoms and ions, depend almost entirely on the feasibility to reach the correct target state, before it can be studied.  We demonstrated how, a \emph{model-free} RL agent can provide valuable insights in constructing variational theories which capture almost all relevant features of the dynamics generated by the optimal protocol. Unlike the optimal bang-bang protocol, the simpler variational protocol is robust to small perturbations, while giving comparable fidelities. This implies the existence of nearly optimal protocols, which do not suffer from the exponential complexity of finding the global minimum of the entire optimization landscape. Finally, in contrast with optimal control methods such as SGD, GRAPE, and CRAB that assume an exact model of the physical system, the model-free nature of RL suggests that it can be used to design protocols even when our knowledge of the physical systems we wish to control is incomplete or our system is noisy or disordered~\cite{bukov2018reinforcement}.

The existence of phase transitions in quantum control problems may have profound consequences beyond physical systems. We suspect that the glassy behavior observed here maybe a generic feature of many control problems and it will be interesting to see if this is indeed the case. It is our hope that given the close connections between optimal control and RL, the physical interpretation of optimization problems in terms of a glassy phase will help in developing novel efficient algorithms and help spur new ideas in RL and artificial intelligence.

\begin{acknowledgments}
We thank J.~Garrahan, M.~Heyl, M.~Schir\'o and D.~Schuster for illuminating discussions. MB, PW and AP were supported by NSF DMR-1813499, ARO W911NF1410540 and AFOSR FA9550-16-1-0334. AD is supported by a NSERC PGS D. AD and PM acknowledge support from Simon's Foundation through the MMLS Fellow program. DS acknowledges support from the FWO as post-doctoral fellow of the Research Foundation -- Flanders and CMTV. We used \href{https://github.com/weinbe58/QuSpin#quspin}{QuSpin} for simulating the dynamics of the qubit systems~\cite{weinberg_17,quspin2}. The authors are pleased to acknowledge that the computational work reported on in this paper was performed on the Shared Computing Cluster which is administered by \href{https://www.bu.edu/tech/support/research/}{Boston University's Research Computing Services}. The authors also acknowledge the Research Computing Services group for providing consulting support which has contributed to the results reported within this paper.
\end{acknowledgments}

\bibliography{ML_bib}

\clearpage

\setcounter{page}{1}
\renewcommand{\thepage}{S\arabic{page}}

\setcounter{table}{0}
\renewcommand{\thetable}{S\arabic{table}}
\renewcommand{\theHtable}{Supplement.\thetable}

\setcounter{figure}{0}
\renewcommand\thefigure{S\arabic{figure}}
\renewcommand{\theHfigure}{Supplement.\thefigure}

\setcounter{equation}{0}
\renewcommand{\theequation}{S\arabic{equation}}
\renewcommand{\theHequation}{Supplement.\theequation}

\setcounter{section}{0}
\renewcommand{\thesection}{S\arabic{section}}
\renewcommand{\theHsection}{Supplement.\thesection}


\onecolumngrid
\begin{center}
	\noindent\textbf{Supplemental Material for:}
	\bigskip
		
	\noindent\textbf{\large{Reinforcement Learning in Different Phases of Quantum Control}}
		
\end{center}

\section{\label{sec:algos}Glassy Behaviour of Different Machine Learning and Optimal Control Algorithms with Local and Nonlocal Flip Updates}

\subsection{\label{subsec:SGD_algo} Stochastic Descent}
To benchmark the results obtained using Reinforcement Learning (RL), we use a greedy stochastic descent (SD) algorithm to sample the infidelity landscape minima containing the driving protocols. We restrict our SD algorithm to exploring bang-bang protocols, for which $h_x(t)\in\{\pm 4\}$. The algorithm starts from a random protocol configuration and proposes local field updates at a time $t$ chosen uniformly in the interval $[0, T]$. The updates consist in changing the applied field $h_x(t)\rightarrow h_x'(t)$ only if this increases the fidelity. Ideally, the protocol is updated until all possible local field updates can only decrease the fidelity. Practically, for some protocol durations, ensuring that a true local minima with respect to $1$-flip is reached can be computational expensive. Therefore, we restrict the number of fidelity evaluations to be at most $20\times T/\delta t$. In this regard, the obtained protocol is a local minimum with respect to local ($1$-flip) field updates. The stochastic descent is repeated multiple times with different initial random protocols. The set of protocols $\{h^\alpha|\alpha=1,\dots,N_\mathrm{real}\}$ obtained with stochastic descent is used to calculate the glass-like order parameter $q(T)$ (see main text). A Python implementation of the algorithm is available on~\href{https://github.com/mgbukov/dynamicQL/tree/master/SA}{Github}.

\subsection{\label{subsec:CRAB} CRAB}

Chopped RAndom Basis (CRAB) is a state-of-the-art optimal control algorithm designed to tackle many-body quantum systems~\cite{doria_11,caneva_11}. The idea behind CRAB is to decompose the unknown driving protocol into a complete basis (Fourier, Laguerre, etc.), and impose a cut-off on the number of `harmonics' kept for the optimisation. The algorithm then uses an optimiser to find the values for the expansion coefficients, which optimise the cost function of the problem. 

Following Ref.~\cite{caneva_11}, we make a Fourier-basis ansatz for the driving protocol.
\begin{eqnarray}
h_\mathrm{CRAB}(t) &=& h_0(t)\left( 1 + \frac{1}{\lambda(t)}\sum_{i=1}^{N_c}A_i\cos\omega_i t + B_i\sin\omega_i t\right),
\end{eqnarray} 
where the Fourier coefficients $\{A_i,B_i,\omega_i\}$ which parametrise the protocol are found using an implementation of the Nelder-Mead optimization method in the SciPy python library. The number of harmonics kept in the optimisation is given by $N_c$. The CRAB algorithm uses two auxiliary functions, defined by the user: the first function $h_0(t)$ is a trial initial guess ansatz for the protocol, while the second function, $\lambda(t)$ imposes the boundary conditions $\lambda\to\infty$ for $t\to 0$ and $t\to T$ to the Fourier expansion term.

The cost function which we optimise in the state manipulation problem
\begin{equation}
\mathcal{C}[h(t)] = \mathcal{F}(\{A_i,B_i,\omega_i\}) + \frac{1}{16T}\int_0^T\mathrm{d}t [h(t)]^2
\end{equation}
contains the fidelity $\mathcal{F}(\{A_i,B_i,\omega_i\})$ at the end of the protocol, and an additional penalty coming from the $L^2$ norm of the protocol to keep the optimal protocols bounded. The last constraint is required for a better and honest comparison with the RL, SD, and GRAPE algorithms.

Applying CRAB to the state manipulation problem from the main text systematically, we choose $\lambda(t)=1/\sin(\pi t/T)^2$, and $h_0(t)=-2+4t/T$. We also consider $N_c=10,20$ to study how much an effect increasing the number of degrees of freedom will have on the optimal protocols found. For each value of $N_c$ we start the optimization algorithm with 10 random initial configurations. We define the optimal protocol as the protocol with the best fidelity out of that group of 10. The random initial frequencies $\omega_i$ are chosen the same way as outlined in Ref.~\cite{caneva_11} while the amplitudes $A_i$ and $B_i$ are chosen uniformly between $-10$ and $10$.

\subsection{\label{subsec:GRAPE} GRAPE}
GRadient Ascend Pulse Engineering, is a numeric derivative-based optimal control method, first introduced in the context of NMR spectroscopy~\cite{grape_05}. As suggested by its name, the method performs gradient optimization. Instead of restricting the protocols to bang-bang type, the method works with quasi-continuous protocols. Protocol magnitudes can take on any value within the allowed manifold but, unlike CRAB, are piecewise constant in time.

In the present case, one can efficiently compute the gradient of the fidelity as follows. Consider the fidelity for some trial protocol $h_x(t)$,
\begin{equation}
F_h(T)=|\left\langle \psi_\ast \right| U(T,0) \left| \psi_i \right\rangle |^2,
\end{equation}
where $U(T,0)$ denotes the time evolution operator from $0$ to $T$. Let us further decompose the Hamiltonian as $H=H_0+h_x(t)X$, where $H_0$ is the part over which we have no control, and $X$ denotes the operator we control. The functional derivative of the fidelity with respect to the protocol thus becomes
\begin{equation}
\frac{\delta F_h(T)}{\delta h_x(t)} = i\left\langle \psi_\ast \right| U(T,0) \left| \psi_i \right\rangle \left\langle \psi_i \right| U(0,t) X U(t,T) \left| \psi_\ast \right\rangle -i \left\langle \psi_\ast \right| U(T,t)XU(t,0) \left| \psi_i \right\rangle \left\langle \psi_i \right| U(0,T) \left| \psi_\ast \right\rangle
\end{equation}
Although this expression appears hard to evaluate, it takes on a very simple form
\begin{equation}
\frac{\delta F_h(T)}{\delta h_x(t)} = 2{\rm Im} \left[\left\langle \phi(t) \right| X \left| \psi_i(t) \right\rangle \right],
\label{eq:grape_grad}
\end{equation}
where $\left| \psi_i(t)\right\rangle=U(t,0)\left| \psi_i\right\rangle$ denotes the initial state propagated to time $t$ and $\left\langle \phi(t) \right|=\left\langle \psi_i(T)| \psi_\ast \right\rangle \left\langle \psi_\ast \right| U(T,t) $ denotes the (scaled) final state propagated back in time to time $t$. Notice that this procedure requires us to exactly know this time evolution operator (i.e. a model of physical system to be controlled). Hence, by propagating both the initial state forward and the target state backward in time one gets access to the full gradient of the control landscape. 

To find a local maximum one can simply now gradient ascend the fidelity. A basic algorithm thus goes as follows:
\begin{enumerate}
	\item[(i)] Pick a random initial magnetic field $h^0_x(t)$.
	\item[(ii)] Compute first $\left| \psi_i(t)\right\rangle$ and then $\left\langle \phi(t) \right|$ for the current setting of the magnetic field $h^{N}_x(t)$.
	\item[(iii)] Update the control field $h^{N+1}_x(t)=h^{N}_x(t)+\epsilon_N {\rm Im} \left[\left\langle \phi(t) \right| X \left| \psi_i(t) \right\rangle \right]$. Note that this step can be upgraded to a second-order Newton method to improve the performance of the algorithm, see Refs.~\cite{machnes2011comparing,agundez_17,de2011second}.
	\item[(iv)] Repeat (ii) and (iii) until the desired tolerance is reached. 
\end{enumerate}
Here $\epsilon_N$ is the step size in each iteration. Choosing a proper step size can be a difficult task. In principle the fidelity should go up after each iteration but if the step size it too large, the algorithm can overshoot the maximum, resulting in a worse fidelity. To avoid this, one should adapt the step size during the algorithm. Numerically we have observed that, in order to avoid overshooting saddles/maxima,  $\epsilon_N \propto 1/\sqrt{N}$. 

\begin{figure}[t!]
	\includegraphics[width=0.5\columnwidth]{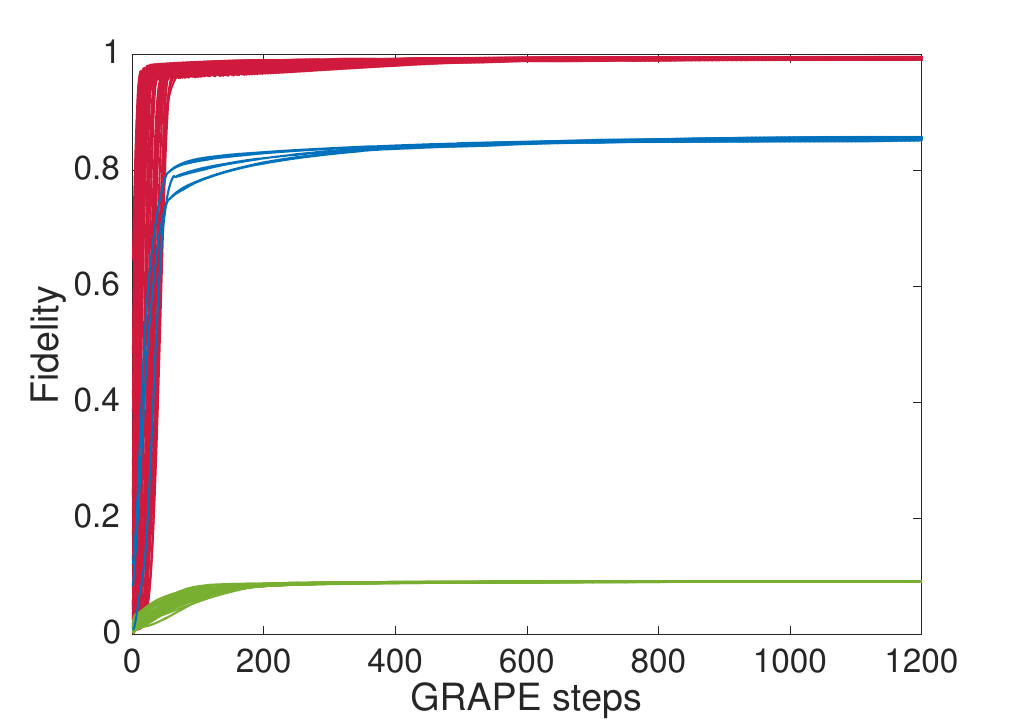}
	\caption{\label{fig:grape_traces}Fidelity traces of GRAPE for $T=3.2$ and $L=6$ as a function of the number of gradient ascend steps for $10^2$ random initial conditions. This figure should be compared to Fig.~\ref{fig:SD_traces} in the main text. GRAPE clearly gets attracted by the same three attractors as SD but has much smaller intra-attractor fluctuations, presumably due to the non-locality of the updates and the continuous values of the control field $h_x\in[-4,4]$ used in GRAPE. This shows that the glass control phase is present for both local and nonlocal update algorithms.}
\end{figure}

\begin{figure}[t!]
	\centering
	\begin{minipage}[b]{0.49\textwidth}
		\includegraphics[width=\textwidth]{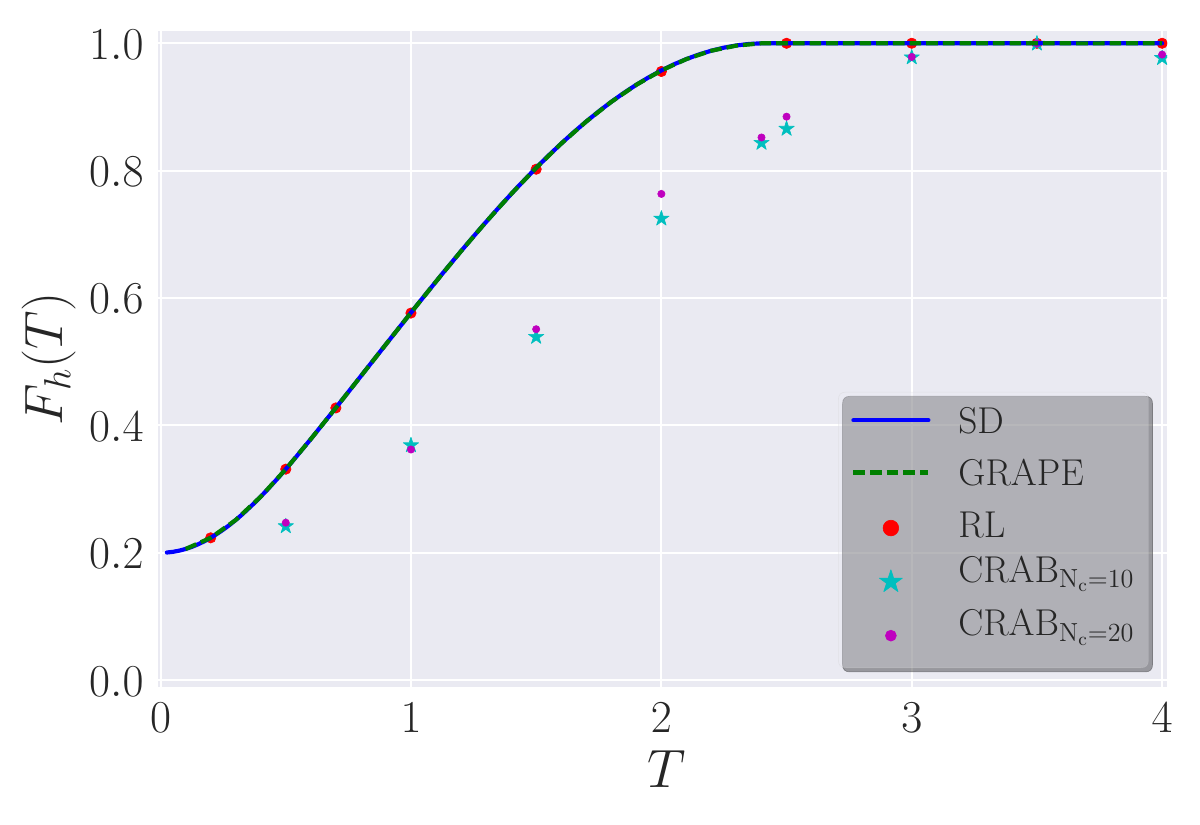}
	\end{minipage}
	\hfill
	\begin{minipage}[b]{0.49\textwidth}
		\includegraphics[width=\textwidth]{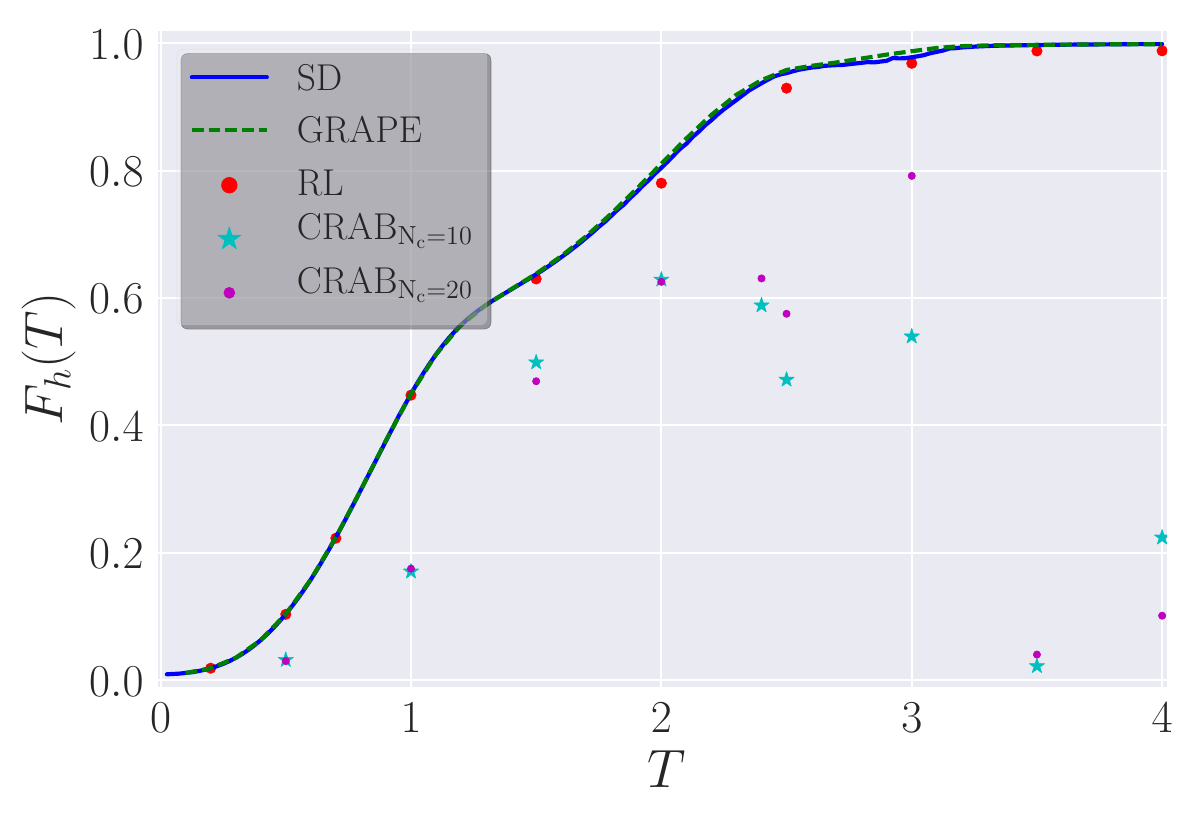}
	\end{minipage}
	\caption{\label{fig:RL_vs_SA} Comparison between the best fidelities obtained using SD (solid blue line), RL (big red dots), GRAPE (dashed green line) and CRAB (cyan star and small magenta dot) for $L=1$ (left) and $L=6$ (right). Here $N_\mathrm{c}$ denotes the cap in the number of harmonics kept in the CRAB simulation.}
\end{figure}

\subsection{\label{subsec:alg_comparison} Comparison between the RL, SD, CRAB and GRAPE}

While a detailed comparison between the ML algorithms and other optimal control algorithms is an interesting topic, it is beyond the scope of the present paper. Below, we only show that both the RL algorithm is capable of finding optimal bang-bang protocols for the quantum control problem from the main text, and that its performance rivals that of SD and the state-of-the-art algorithms for many-body quantum problems CRAB and GRAPE.

The result for the single qubit $L=1$ are shown in Fig.~\ref{fig:RL_vs_SA} (left). The most important points can be summarised as follows:
\begin{itemize}
	\item RL, GS and GRAPE all find the optimal protocol.
	\item Below the quantum speed limit, $T_\mathrm{QSL}$, CRAB finds good, but clearly suboptimal protocols. The plots also show the glassiness represents a generic feature of the constrained optimization problem and not the method used to perform the optimization. Increasing the cutoff $N_c$, and with it the number of effective degrees of freedom, does not lead to a sizeable improvement in CRAB. One explanation is the following: for the single qubit, we know that the variational protocol, which contains at most two bangs (see Fig.~\ref{fig:phase_diag}b), is a global minimum of the optimisation landscape. Such a protocol can easily be approximated using up to $N_c=20$ harmonics. However, allowing more degrees of freedom comes at a huge cost due to the glassiness of the problem: there exist many quasi-degenerate local minima for the algorithm to get stuck in.
\end{itemize}

The comparison for the Many Coupled Qubits system for $L=6$ are shown in Fig.~\ref{fig:RL_vs_SA} (right). Larger values of $L$ do not introduce any change in the behaviour, as we argue in a subsequent section below. In the many-body case, the variational protocol, which contains four bangs, is shown \emph{not} to be the global minimum of the infidelity landscape. Instead, the true global minimum contains many more bangs which, however, only marginally improve the fidelity.
\begin{itemize}
	\item All algorithms give reasonable fidelities, see Fig.~\ref{fig:RL_vs_SA} (right)
	\item Even though GRAPE seems to display better performance out of the four methods, one should not forget that this algorithm, which uses global flips, requires knowledge of all fidelity gradients -- valuable information which is not easily accessible through experimental measurements. One has to keep in mind though, that this comparison is not completely honest, since GRAPE allows for the control field $h_x$ to take any value in the interval $[-4,4]$, which offers a further advantage over the bang-bang based RL. On the other hand, the model-free RL rivals the performance of GRAPE at all protocol durations, and outperforms CRABS, even for the single qubit below the quantum speed limit, where the problem enters the glassy phase. The reason for the seemingly slowly decreasing performance of RL with $T$ is that, since the total protocol duration is held fixed, the number of bangs increases exponentially with $T\sim N_T$, and hence the state space which has to be explored by the agent also grows exponentially. A detailed study of the scaling is postponed to future studies.
	\item All algorithms suffer from the glassiness in the optimisation landscape. This is not surprising, since the glass phase is an intrinsic property of the infidelity landscape, as defined by the optimisation problem, and does not depend on which algorithm is used to look for the optimal solution. Fig.~\ref{fig:grape_traces} shows the fidelity traces as a function of running time for GRAPE. Comparing this to the corresponding results for SD, see Fig.~\ref{fig:SD_traces}, we see a strikingly similar behaviour, even though GRAPE uses nonlocal flip updates in contrast to SD. This means that, in the glassy phase, GRAPE also gets stuck in suboptimal attractors, similar to SD and RL. Thus, as an important consequence, the glassy phase affects both local and nonlocal-update algorithms. 
\end{itemize}

\section{\label{sec:2LS_comparison}Performance of the Different Driving Protocols for the Qubit}

\begin{figure*}[t!]	
	\includegraphics[width=1.0\columnwidth]{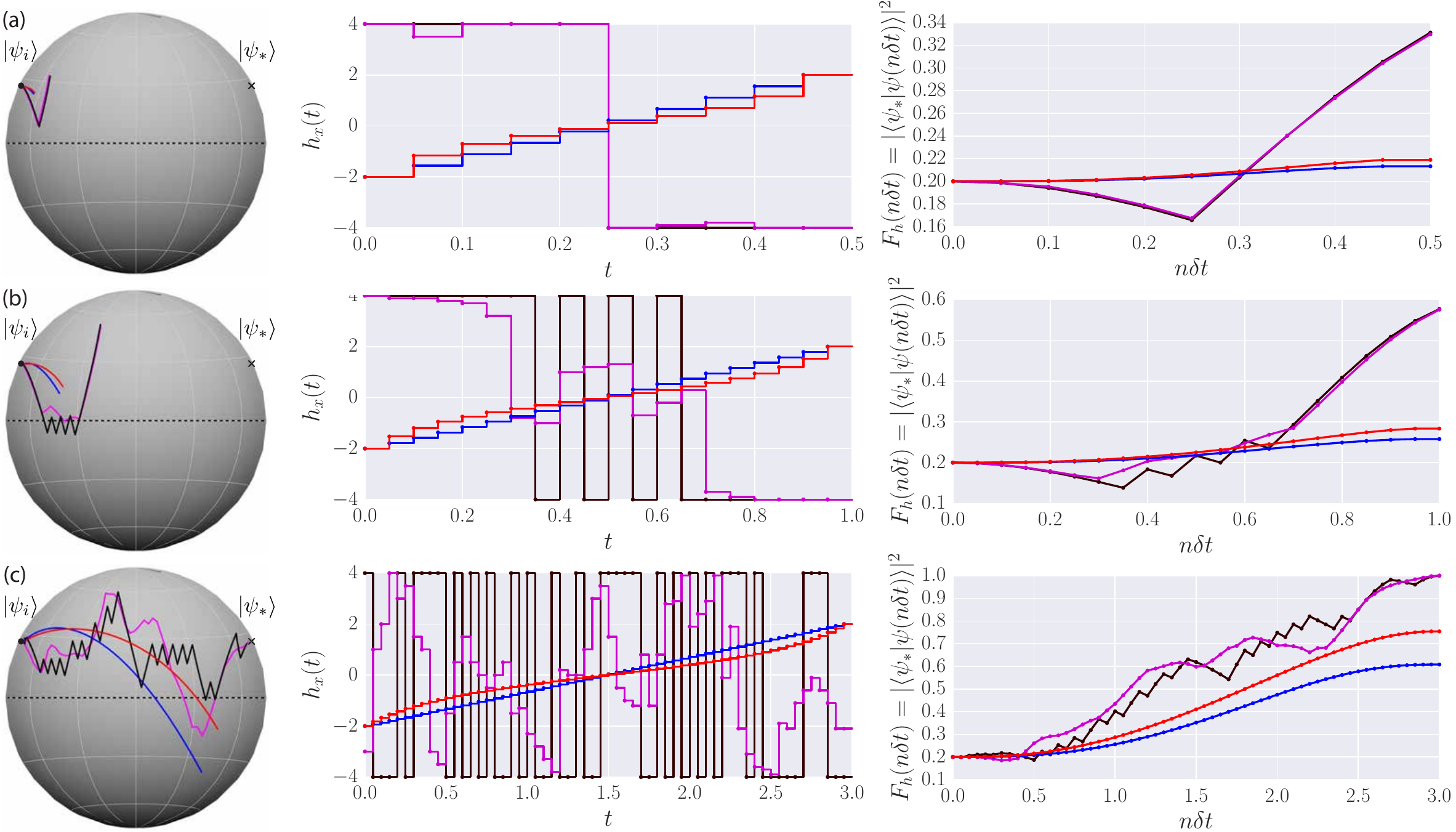}
	\caption{\label{fig:comapres_protocols} Comparison between the bang-bang (black) and quasi-continuous (magenta) protocols found by the RL agent, and the Landau-Zener (blue) and geodesic (red) protocols computed from analytical theory in the overconstrained phase for $T=0.5$ (a), the glassy phase for $T=1.0$ (b), and the controllable phase for $T=3.0$ (c). The left column shows the representation of the corresponding protocol on the Bloch sphere, the middle one -- the protocols themselves, and the right column -- the instantaneous fidelity in the target state $\psi_\ast\rangle$.}
\end{figure*}

It is interesting to compare the bang-bang and quasi-continuous driving protocols found by the agent to a simple linear protocol, which we refer to as Landau-Zener (LZ), and the geodesic protocol, which optimizes local fidelity close to the adiabatic limit essentially slowing down near the minimum gap~\cite{tomka_16}. We find that the RL agent offers significantly better solutions in the overconstrained and glassy phases, where the optimal fidelity is always smaller than unity. The Hamiltonian of the qubit together with the initial and target states read:
\begin{equation}
H(t)=-S^z-h_x(t)S^x,\qquad |\psi_i\rangle\sim(-1/2 - \sqrt{5}/2, 1)^{\mathsf{T}}, \qquad |\psi_\ast\rangle\sim(1/2 + \sqrt{5}/2, 1)^{\mathsf{T}},
\end{equation}
where $|\psi_i\rangle$ and $|\psi_\ast\rangle$ are the ground state of $H(t)$ for $h_i = -2$ and $h_\ast = +2$ respectively. Note that for bang-bang protocols, the initial and target states are not eigenstates of the control Hamiltonian since $h_x(t)$ takes on the values $\pm 4$.

The RL agent is initiated at the field $h(t=0)=h_\mathrm{min}=-4.0$. The RL protocols are constructed from the following set of jumps, $\delta h_x$, allowed at each protocol time step $\delta t$:
\begin{itemize}
	\item \emph{bang-bang} protocol: $\delta h_x\in\{0.0,\pm 8.0\}$ which, together with the initial condition, constrains the field to take the values $h_x(t)\in\{\pm 4.0 \}$.
	\item \emph{quasi-continuous} protocol: $\delta h_x\in\{ 0.0,\pm 0.1,\pm 0.2, \pm 0.5, \pm 1.0, \pm 2.0, \pm 4.0, \pm 8.0 \}$. We restrict the actions available in a state to ensure $h_x(t)\in[-4.0,4.0]$.
\end{itemize}
Interestingly, the RL agent figures out that it is always advantageous to first jump to $h_\mathrm{max}=+4.0$ before starting the evolution, as a consequence of the positive value of the coefficient in front of $S^z$.

The analytical adiabatic protocols are required to start and end in the initial and target states, which coincide with the ground states of the Hamiltonians with fields $h_i=-2.0$ and $h_\ast=2.0$, respectively. They are defined as follows:
\begin{itemize}
	\item \emph{Landau-Zener(LZ)} protocol: $h_x(t)=(h_\ast-h_i)t/T + h_i$
	\item \emph{geodesic} protocol: $h_x(t)=\tan(at + b)$, where $b=\arctan(h_i)$ and $a=\arctan(h_\ast-b)/T$.
\end{itemize}

Figure~\ref{fig:comapres_protocols} shows a comparison between these four protocol types for different values of $T$, corresponding to the three quantum control phases. Due to the instantaneous gap remaining small compared to the total protocol duration, the LZ and geodesic protocols are very similar, irrespective of $T$. The two protocols significantly differ only at large $T$, where the geodesic protocol significantly outperforms the linear one.  An interesting general feature for the short protocol durations considered is that the fidelities obtained by the LZ and geodesic protocols are clearly worse than the ones found by the RL agent. This points out the far-from-optimal character of these two approaches, which essentially reward staying close to the instantaneous ground state during time evolution. Looking at the fidelity curves in Fig.~\ref{fig:comapres_protocols}, we note that, before reaching the optimal fidelity at the end of the ramp for the overconstrained and glassy phases, the instantaneous fidelity drops below its initial value at intermediate times. This suggests that the angle between the initial and target states on the Bloch sphere becomes larger in the process of evolution, before it can be reduced again. Such situation is very reminiscent of counter-diabatic or fast forward driving protocols, where the system can significantly deviate from the instantaneous ground state at intermediate times~\cite{sels_16,zhou_17,jarzynski_17}. Such problems, where the RL agent learns to sacrifice local rewards in view of obtaining a better total reward in the end are of particular interest in RL~\cite{sutton_barto}.

\section{\label{sec:critical_scaling_2LS}Critical Scaling Analysis of the Control Phase Transitions}

\begin{figure}[t!]
	\centering
	\begin{subfigure}[b]{0.496\textwidth}
		\includegraphics[width=\textwidth]{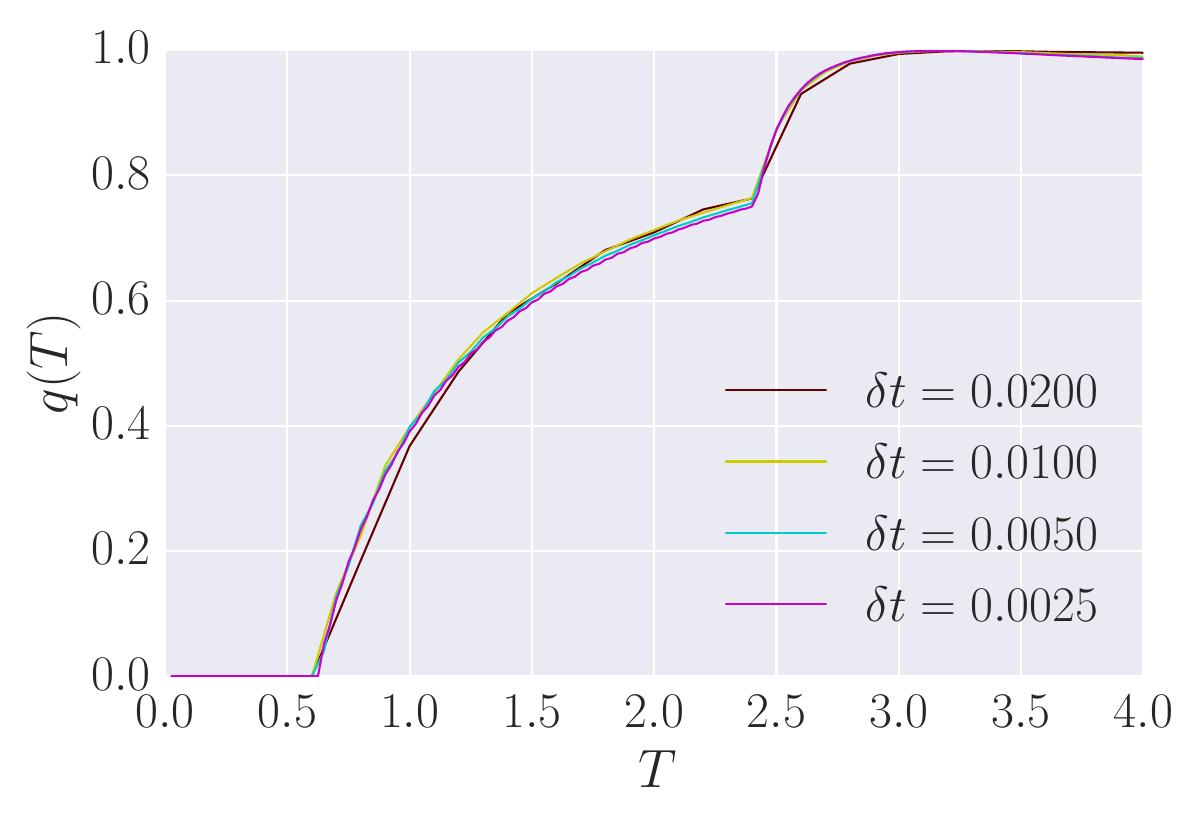}
		\caption{$L=1$}
	\end{subfigure}
	\begin{subfigure}[b]{0.496\textwidth}
		\includegraphics[width=\textwidth]{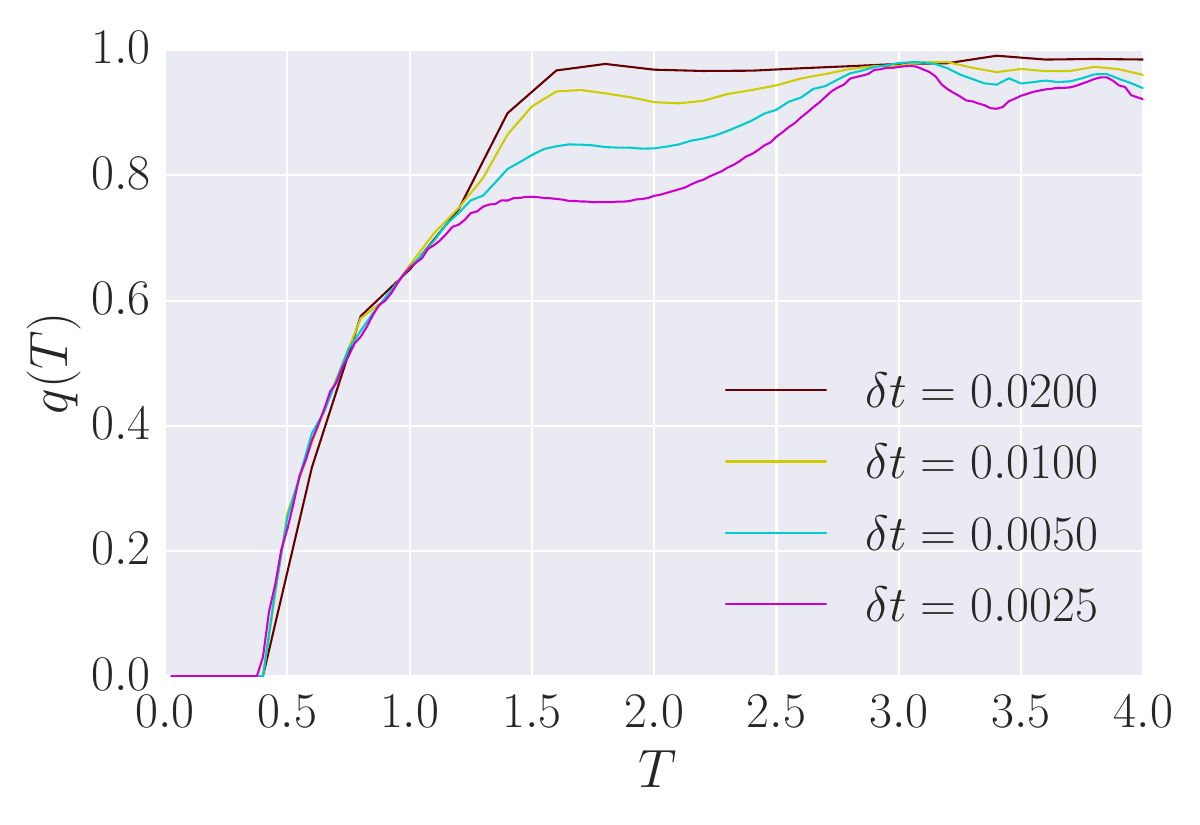}
		\caption{$L=10$}
	\end{subfigure}
	\begin{subfigure}[b]{0.496\textwidth}
		\includegraphics[width=\textwidth]{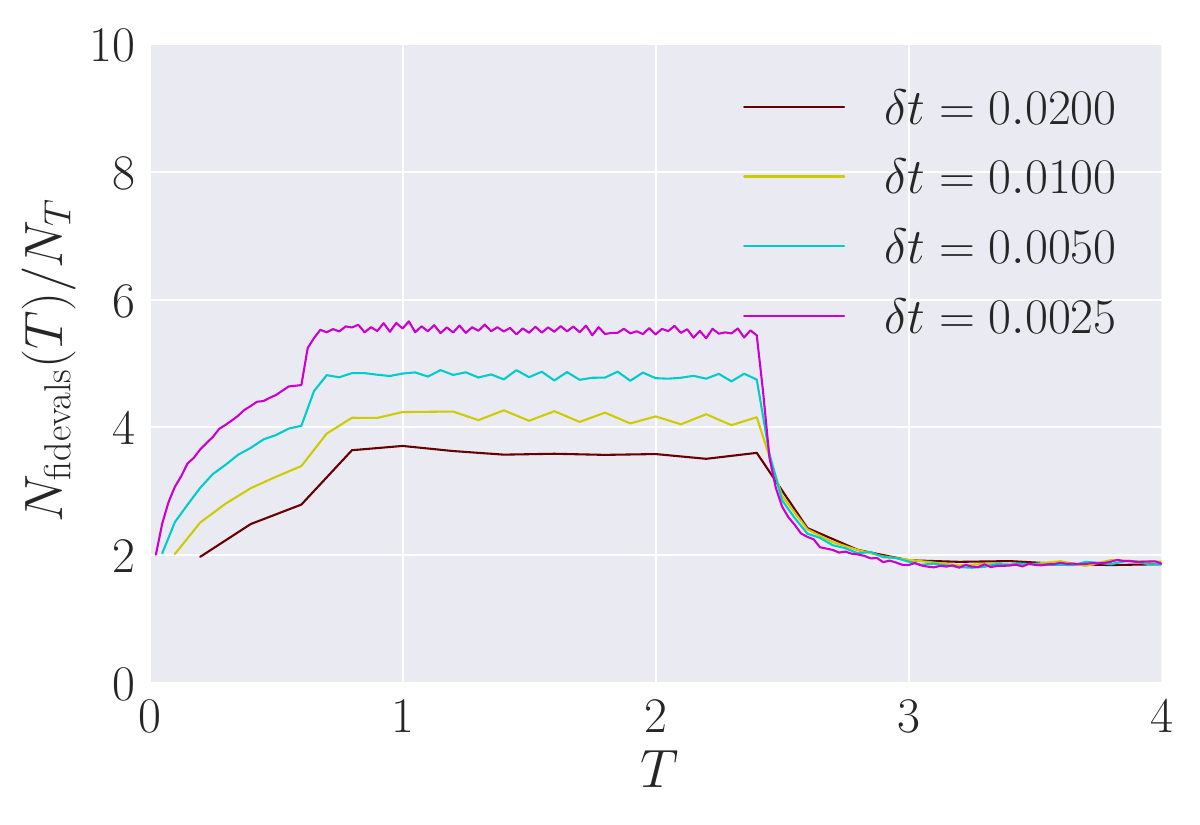}
		\caption{$L=1$}
	\end{subfigure}
	\begin{subfigure}[b]{0.496\textwidth}
		\includegraphics[width=\textwidth]{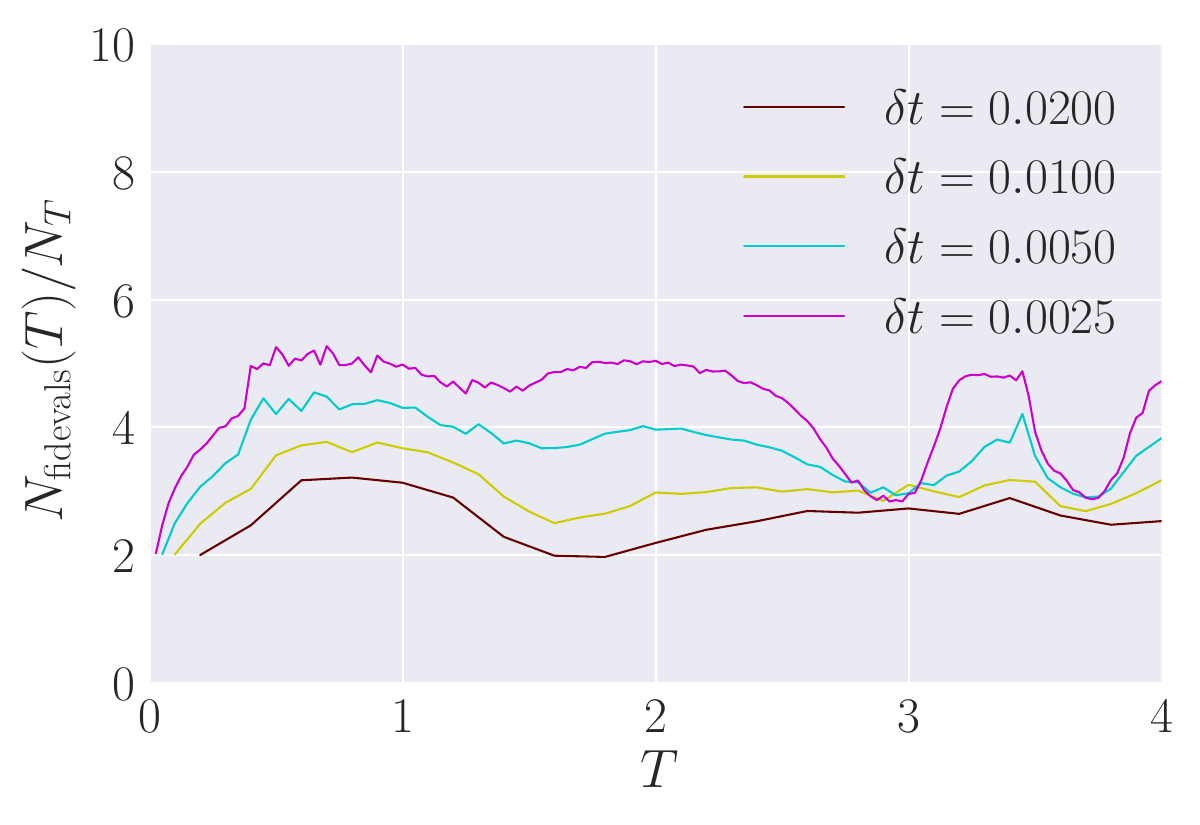}
		\caption{$L=10$}
	\end{subfigure}
	\caption{\label{fig:scaling_dt} Finite time step-size, $\delta t$, scaling of the order parameter $q(T)$ [top] and the number of fidelity evaluations per time step, $N_\mathrm{fidevals}/N_T$ [bottom], to reach a local minimum of the infidelity landscape.}  
\end{figure} 

In this section, we show convincing evidence for the existence of the phase transitions in the quantum state manipulation problem discussed in the main text. We already argued that there is a phase transition in the optimization problem as a function of the protocol time $T$. Mathematically the problem is formulated as
\begin{equation}
h^\mathrm{optimal}_x(t) = \argmin_{h_x(t): |h_x(t)|\leq 4}\{I_h(T)\} = \argmax_{h_x(t): |h_x(t)|\leq 4}|\langle\psi_\ast|\mathcal{T}_t \mathrm e^{-i\int_0^T\mathrm{d}t H[h_x(t)]}|\psi_i\rangle|^2,
\end{equation}
i.e.~as finding the optimal driving protocol $h^\mathrm{optimal}_x(t)$, which transforms the initial state (the ground state of the Hamiltonian $H$ at $h_x=-2$) into the target state (the ground state of $H$ corresponding to $h_x=2$) maximizing the fidelity in protocol duration $T$ under unitary Schr\"odinger evolution. We assume that the ramp protocol is bounded, $h_x(t)\in[-4,4]$, for all times during the ramp. In this section, we restrict the analysis to bang-bang protocols only, for which $h_x(t)\in\{\pm 4\}$. The minimum protocol time step is denoted by $\delta t$. There are two different scaling limits in the problem. We define a continuum limit for the problem as $\delta t\to 0$ while keeping the total protocol duration $T=const$. Additionally, there is the conventional thermodynamic limit, where we send the system size $L\to\infty$.

As we already alluded to in the main text, one can think of this optimization problem as a minimization in the infidelity landscape, determined by the mapping $h_x(t)\mapsto I_h(T)=1-F_h(T)$, where each protocol is assigned a point in fidelity space -- the probability of being in the target state after evolution for a fixed protocol duration $T$. Finding the global minimum of the landscape then corresponds to obtaining the optimal driving protocol for any fixed $T$.

To obtain the set of local minima $\{h_x^\alpha(t)|\alpha=1,\dots,N_\mathrm{real}\}$ of the infidelity landscape at a fixed total protocol duration $T$ and protocol step size $\delta t$, we apply Stochastic Descent(SD), see above, starting from a random protocol configuration, and introduce random \emph{local} changes to the bang-bang protocol shape until the fidelity can no longer be improved. This method is guaranteed to find a set of representative local infidelity minima with respect to ``$1$-flip" dynamics, mapping out the bottom of the landscape of $I_h(T)$. Keeping track of the mean number of fidelity evaluations $N_\mathrm{fidevals}$ required for this procedure, we obtain a measure for the average time it takes the SD algorithm to settle in a local minimum. While the order parameter $q(T)$ (see below) was used in the main text as a measure for the static properties of the infidelity landscape, dynamic features are revealed by studying the number of fidelity evaluations $N_\mathrm{fidevals}$. 

As discussed in the main text, the rich phase diagram of the problem can also be studied by looking at the order parameter function $q$ (closely related to the Edwards-Anderson order parameter for detecting glassy order in spin systems~\cite{castellani_05}):
\begin{equation}
q(T)=\frac{1}{16N_T}\overline{\sum_{n=1}^{N_T}\{h_x(n\delta t)-\overline{h_x}(n\delta t)\}^2} ,\qquad \qquad \overline{h_x}(t)=\frac{1}{N_\mathrm{real}}\sum_{\alpha=1}^{N_\mathrm{real}}h_x^\alpha(t).
\end{equation}
Here, $N_T$ is the total number of protocol time steps of fixed width $\delta t$, $N_\mathrm{real}$ is the total number of random protocol realisations $h_x^\alpha(t)$ probing the minima of the infidelity landscape (see previous paragraph), and the factor $1/16$ serves to normalise the squared bang-bang drive protocol $h_x^2(t)$ within the range $[-1,1]$.

\subsection{Single Qubit}

For $T>T_\mathrm{QSL}$, the optimization problem of the single qubit ($L=1$) is controllable, and there exist infinitely many protocols which can prepare the target state with unit fidelity. In analogy with the random $k$-SAT problem~\cite{mezard_02},
we call this the \emph{controllable} (or underconstrained) phase of the quantum control problem. Intuitively, this comes about due to the large total protocol durations available which allow one to correct a set of `wrong moves' at a later time in order to achieve a better fidelity at the end of the ramp. We have seen that both the Reinforcement Learning (RL) and Stochastic Descent (SD) agents readily and quickly learn to exploit this feature for optimal performance. In this phase, which is not present in the thermodynamic limit $L\to\infty$, there exist analytical results to compute driving protocols of unit fidelity based on the geodesic and counter-diabatic approaches~\cite{hegerfeldt_13,kolodrubetz_16,sels_16}. The driving protocols $h_x^\alpha(t)$, corresponding to the minima of the infidelity landscape $I_h(T)$, are completely uncorrelated, resulting in $\overline{h_x}(t)=0$ and, thus, $q=1$.
As $T\searrow T_\mathrm{QSL}$, the infidelity minima start becoming correlated, reflected in a drop in the value of $q(T)$. At the time $T=T_\mathrm{QSL}$, a phase transition occurs to a glassy phase with shallow, quasi-degenerate infidelity minima corresponding to many almost equally optimal protocols. Fig.~\ref{fig:scaling_dt}a shows that the order parameter $q(T)$ develops a clear non-analyticity in the continuum limit $\delta t\to 0$, which proves the existence of a phase transition in protocol space. At the same critical time $T_\mathrm{QSL}$, a drastic rapid increase is detected in the number of fidelity evaluations required to map out the infidelity minima, see Fig.~\ref{fig:scaling_dt}c.

\begin{figure}[t!]
	\centering
	\begin{subfigure}[b]{0.496\textwidth}
		\includegraphics[width=\textwidth]{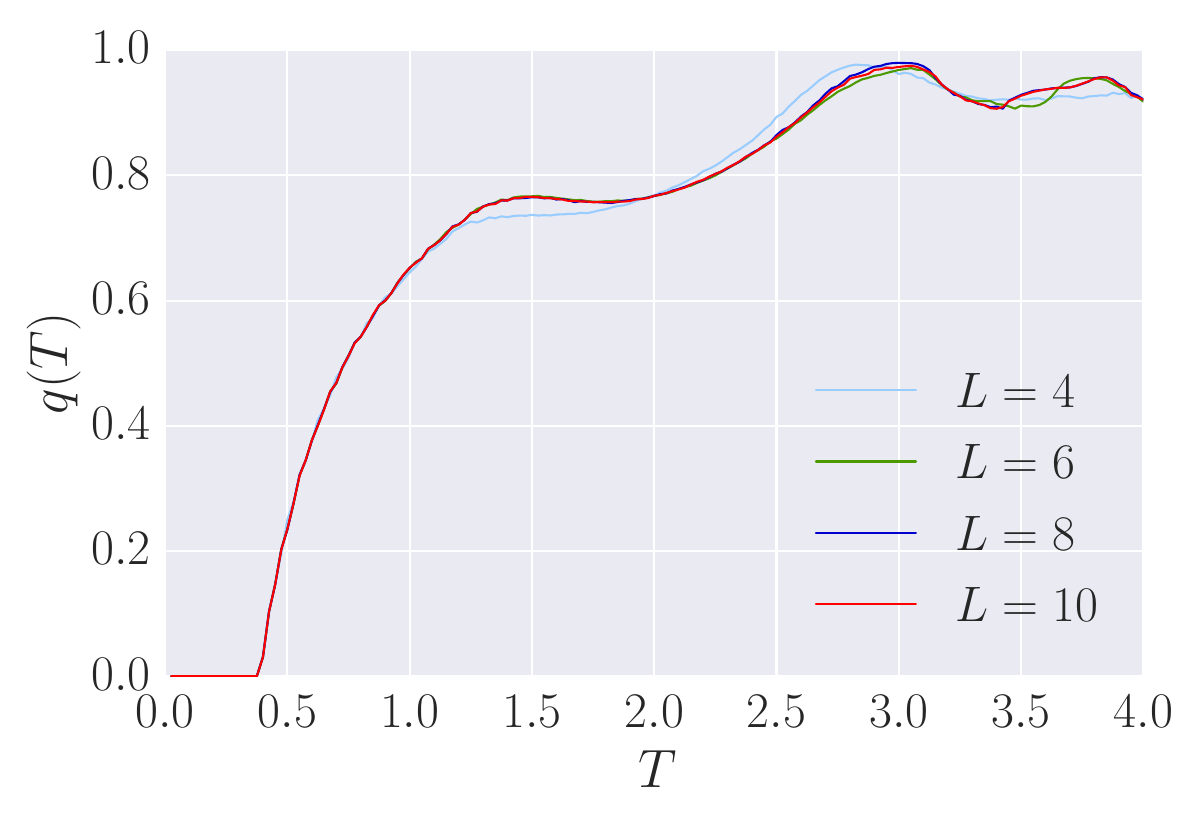}
		\caption{Order parameter $q(T)$ vs.~total ramp duration $T$.}
	\end{subfigure}
	\begin{subfigure}[b]{0.496\textwidth}
		\includegraphics[width=\textwidth]{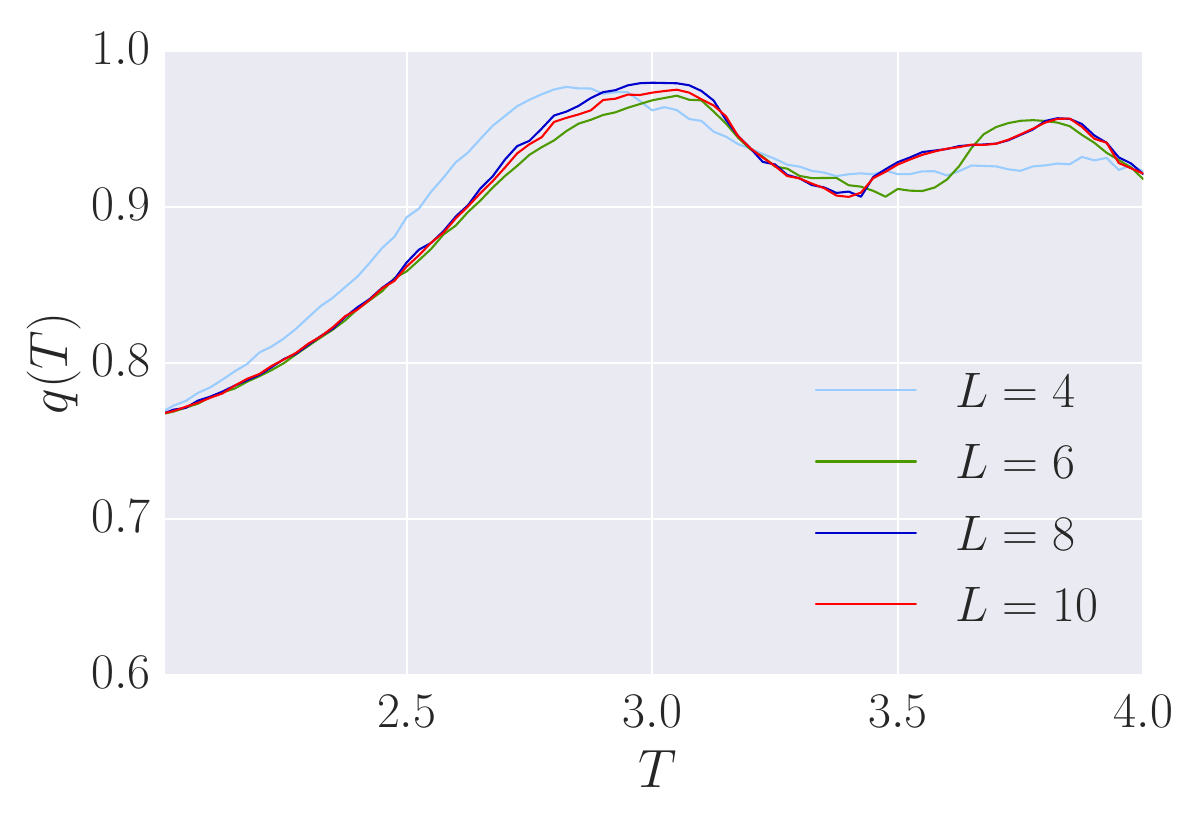}
		\caption{same as (a) with later times zoomed in.}
	\end{subfigure}
	\begin{subfigure}[b]{0.496\textwidth}
		\includegraphics[width=\textwidth]{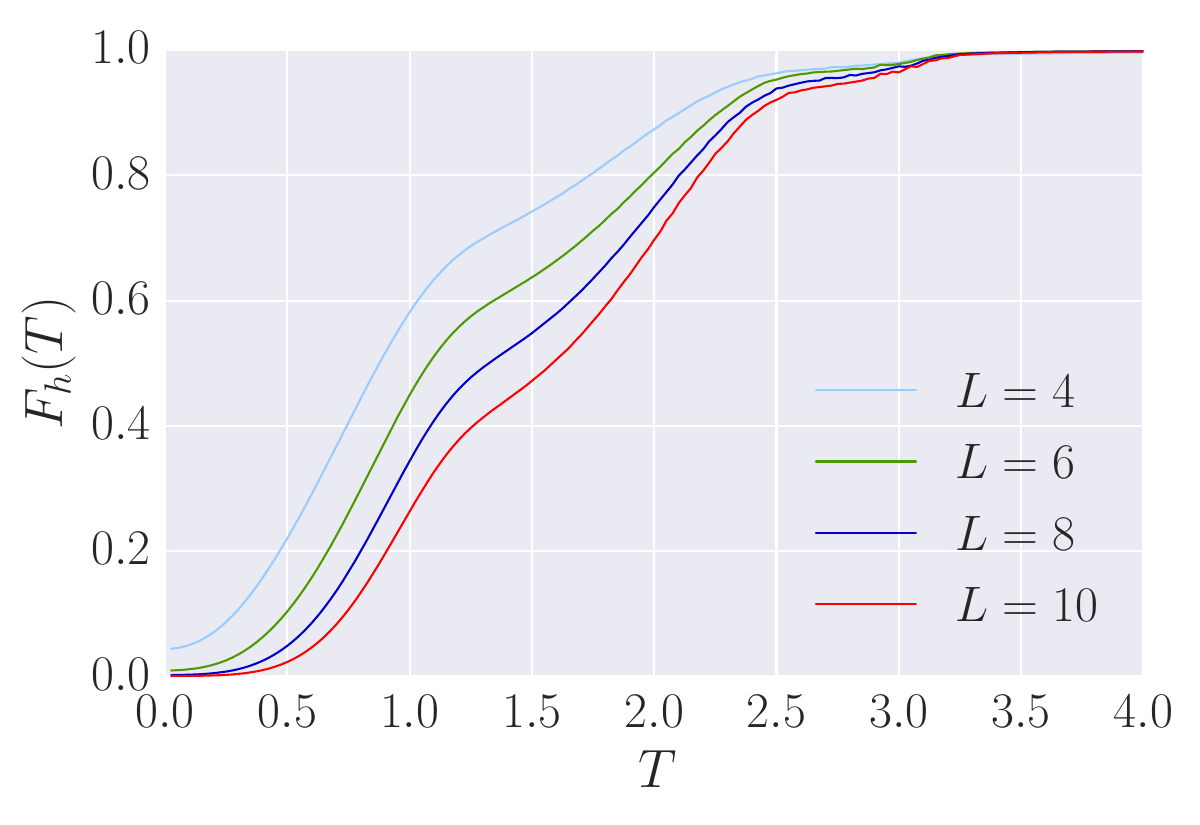}
		\caption{Optimal protocol many-body fidelity: linear scale. Convergence is reached at $L\geq 6$.}
	\end{subfigure}
	\begin{subfigure}[b]{0.496\textwidth}
		\includegraphics[width=\textwidth]{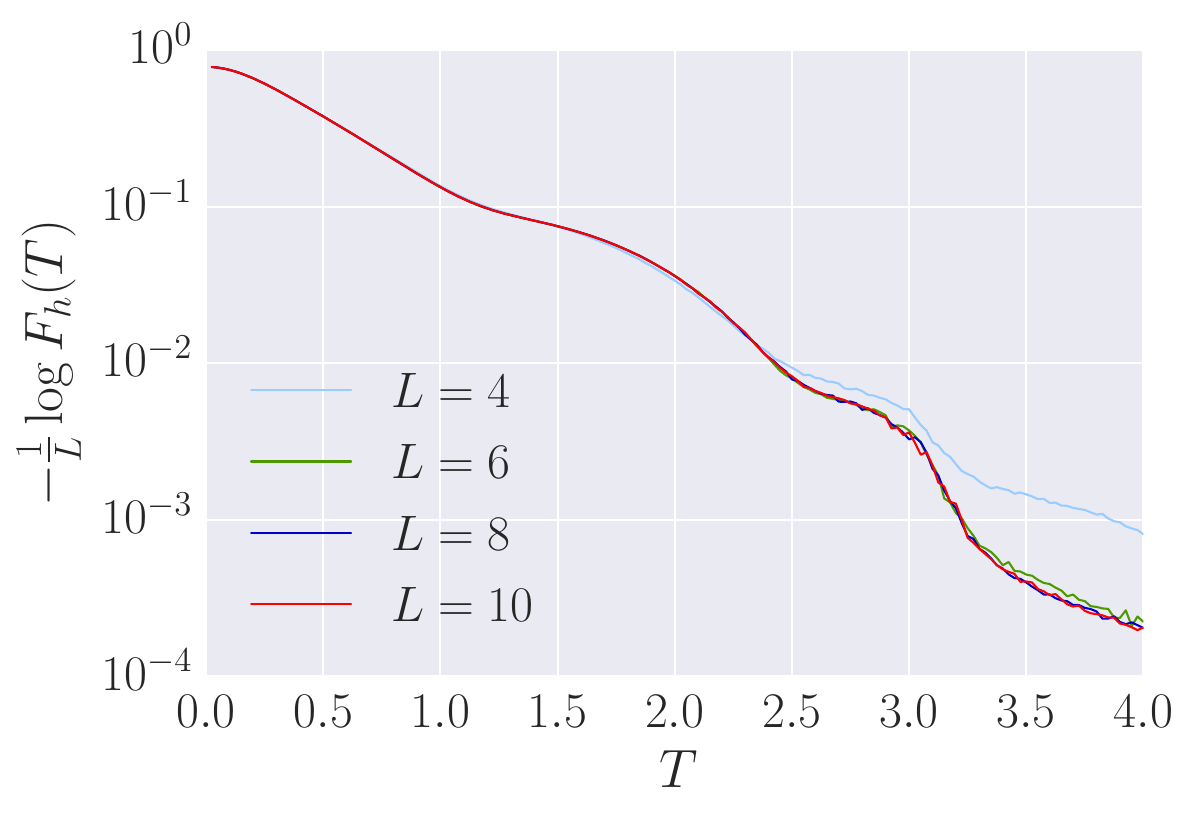}
		\caption{Optimal protocol many-body fidelity: logarithmic scale. Convergence is reached at $L\geq 6$.}
	\end{subfigure}
	\begin{subfigure}[b]{0.496\textwidth}
		\includegraphics[width=\textwidth]{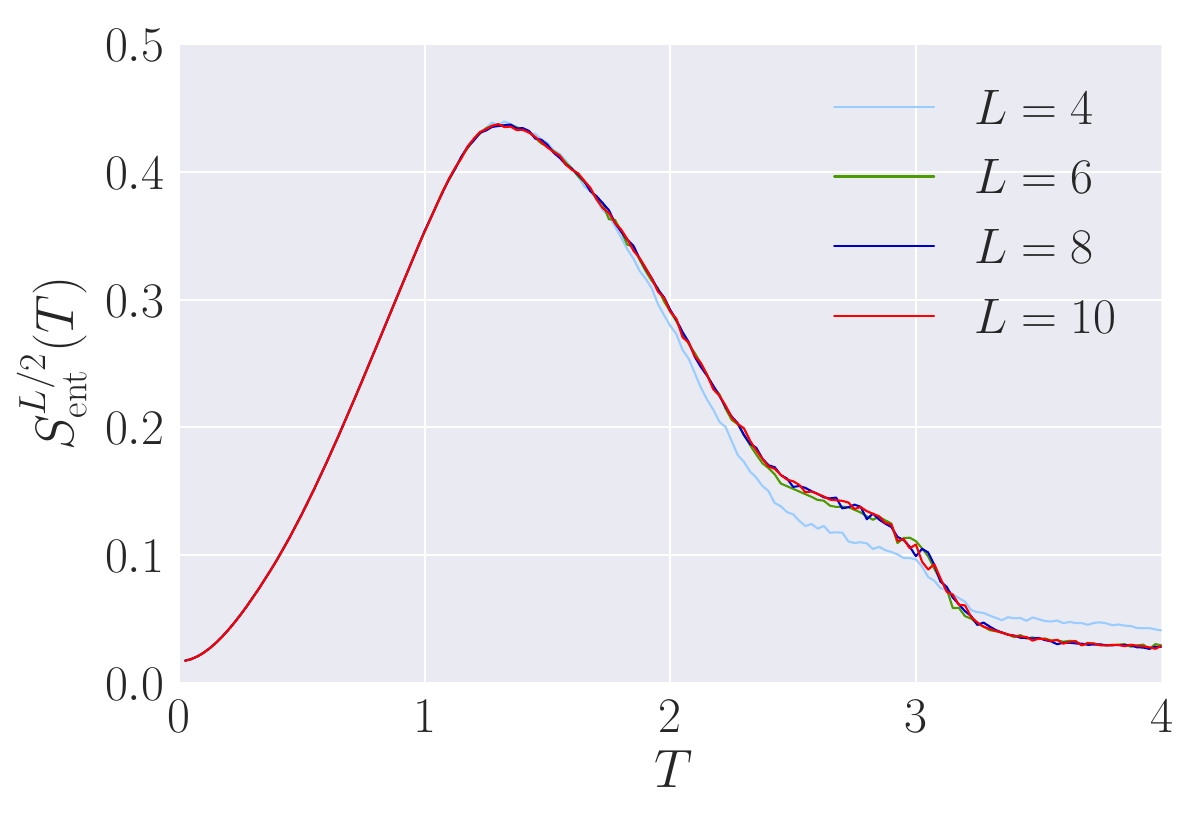}
		\caption{Entanglement entropy of the half chain as a function of the system size $L$.}
	\end{subfigure}
	\begin{subfigure}[b]{0.496\textwidth}
		\includegraphics[width=\textwidth]{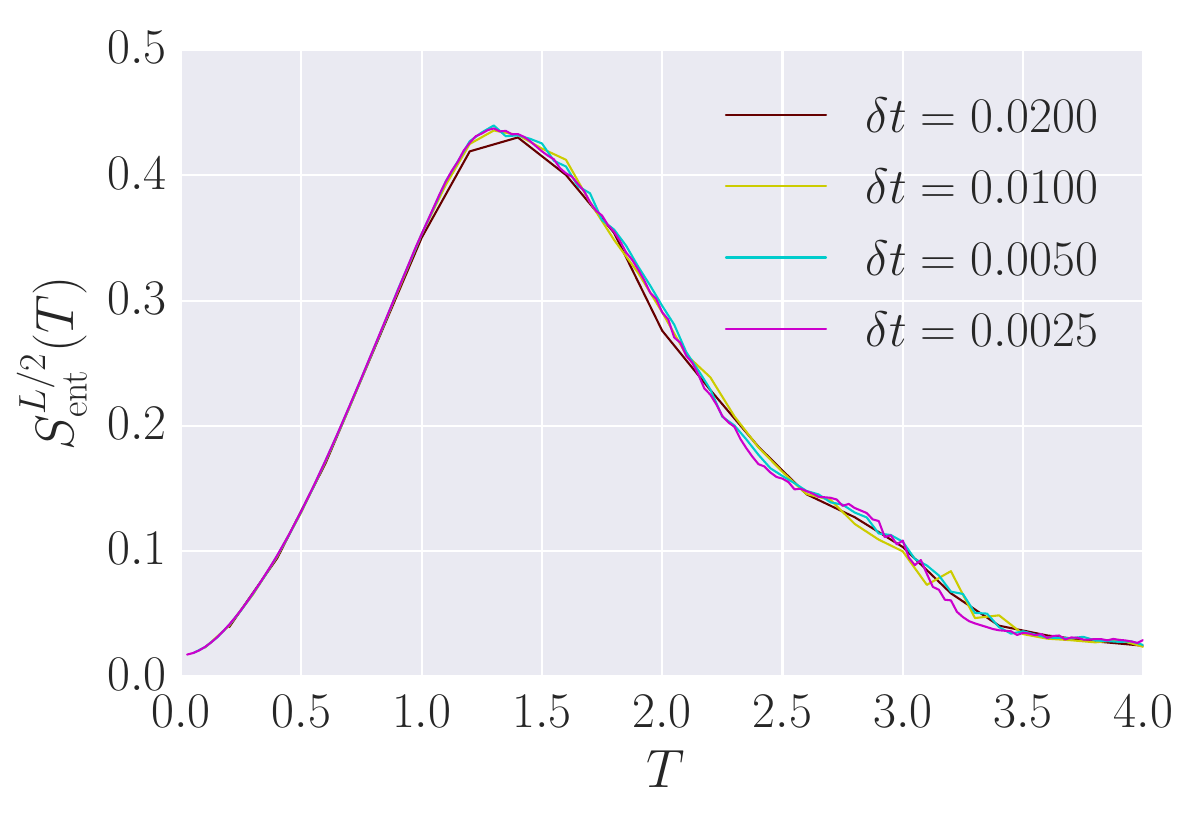}
		\caption{Entanglement entropy of the half chain as a function of the protocol step size $\delta t$, for $L=10$.}
	\end{subfigure}
	\caption{\label{fig:scaling_L} Finite system-size $L$ scaling of the order parameter $q(T)$[top], the many-body fidelity $F_h(T)$ [middle] and the entanglement entropy $S_\mathrm{ent}^{L/2}$ for protocol step size $\delta t=0.0025$.}  
\end{figure} 

For $T_c<T<T_\mathrm{QSL}$ the control problem is in the glassy phase. We showed in the main text by examining the DOS of protocols with respect to local flips of the field, that finding the optimal protocol appears as complicated as finding the ground state of a glass. This is reflected in the observed increase of the average number of fidelity evaluations $N_\mathrm{fidevals}$ with decreasing $\delta t$ (c.f~Fig.~\ref{fig:scaling_dt}c), and a decrease in the learning rate of the RL and SD agents. The local minima protocols $\{h_x^\alpha(t)\}$ are strongly correlated, as can be seen from the finite value of the order parameter $q(T)$ in Fig.~\ref{fig:scaling_L}a. More importantly, for any practical purposes, unit fidelity can no longer be obtained under the given dynamical constraints.

When we reach the second critical point $T=T_c$, another phase transition occurs from the glassy to an overconstrained phase. At $T=T_c$, the order parameter reaches zero, suggesting that the infidelity landscape contains a single minimum. In this phase, i.e.~for $T<T_c$, the protocol duration is too short to achieve a good fidelity. Nonetheless, in the continuum limit $\delta t\to 0$, there exists a single optimal protocol, although the corresponding maximum fidelity is far from unity. In this overconstrained phase, the optimization problem becomes convex and easy to solve. This is reflected by the observation both the optimal quasi-continuous and bang-band protocols found by the RL agent are nearly identical, cf.~Fig.~\ref{fig:comapres_protocols}. The dynamic character of the phase transition is revealed by a sudden drop in the number of fidelity evaluations $N_\mathrm{fidevals}$.

\subsection{Coupled Qubits}

One can also ask the question what happens to the quantum control phases in the thermodynamic limit, $L\to\infty$. To this end, we repeat the calculation for a series of chain lengths $L$. We omit the case $L=2$, in which the physics of the control problem has a different character, exhibiting spontaneous symmetry breaking in the glasy phase~\cite{bukov_17symmbreak}. Due to the non-integrable character of the many-body problem, we are limited to small system sizes. However, Fig.~\ref{fig:scaling_L} shows convincing data that we capture the behaviour of the system in the limit $L\to\infty$ for the relatively short protocol durations under consideration. Moreover to our surprise the finite-size effects almost entirely disappear for $L\geq 6$ for all range of protocol durations we are considering. It seems that system is able to find an optimal solution, where the information simply does not propagate outside of a very small region and hence the optimal protocol rapidly becomes completely insensitive to the system size.

Figure~\ref{fig:scaling_L}c-d shows the system size scaling of the negative logarithmic \emph{many-body} fidelity. While at $L=4$ we do see remnants of finite-size effects, starting from $L=6$ the curves are barely changing.
A similar rapid system-size convergence is observed also for the order-parameter $q(T)$ (see Fig.~\ref{fig:scaling_L}a-b) and the entanglement entropy of the half chain (Fig.~\ref{fig:scaling_L}e). The protocol step size dependence of the order parameter $q(T)$, the average fidelity evaluations $N_\mathrm{fidevals}$, and the entanglement entropy $S_\mathrm{ent}^{L/2}$ of the half-chain are shown in Figs.~\ref{fig:scaling_dt}b,~\ref{fig:scaling_dt}d and~\ref{fig:scaling_L}f.

\section{Video Material}

This paper is accompanied by eight short videos, labeled Video $1$ through $8$. The videos are available as part of the Supplemental Material published together with the paper. A legend for the videos can be found on \href{https://mgbukov.github.io/RL_movies/}{https://mgbukov.github.io/RL\_movies/}.

\end{document}